\documentclass[]{iopart}      
\usepackage{iopams}
\usepackage{color}
\usepackage{graphicx}

\def\hrss{h_\mathrm{rss}}
\def\hrssdet{h_\mathrm{det}^\mathrm{rss}}
\def\hrssmid{h_\mathrm{rss}^\mathrm{mid}}
\def\Zg{Z_\mathrm{g}}

\def\dccnumber{LIGO-P060016-C-Z}

\begin{document}

\title[Search for gravitational-wave bursts in LIGO data]
      {Search for gravitational-wave bursts in LIGO data from the fourth science run}

\author{B~Abbott$^{14}$, R~Abbott$^{14}$, R~Adhikari$^{14}$, J~Agresti$^{14}$, P~Ajith$^{2}$,
B~Allen$^{2,51}$, R~Amin$^{18}$, S~B~Anderson$^{14}$, W~G~Anderson$^{51}$, M~Arain$^{39}$
% ...
\\ (LIGO Scientific Collaboration)}
% ...
\address{$^{2}$ Albert-Einstein-Institut, Max-Planck-Institut f\"ur Gravitationsphysik, D-30167 Hannover, Germany}
% ...
\address{$^{14}$ LIGO - California Institute of Technology, Pasadena, CA  91125, USA}
% ...
\address{$^{18}$ Louisiana State University, Baton Rouge, LA  70803, USA}
% ...
\address{$^{39}$ University of Florida, Gainesville, FL  32611, USA}
% ...
\address{$^{51}$ University of Wisconsin-Milwaukee, Milwaukee, WI  53201, USA}
% ...

\ead{\mailto{pshawhan@umd.edu}}

\begin{abstract}
The fourth science run of the LIGO and GEO 600 gravitational-wave detectors,
carried out in early 2005, collected data with significantly lower
noise than previous science runs.  We report on a search for
short-duration gravitational-wave bursts with arbitrary waveform in
the $64$--$1600$~Hz frequency range appearing in all three LIGO
interferometers.  Signal consistency tests, data quality cuts,
and auxiliary-channel vetoes are applied to reduce the rate of
spurious triggers.  No gravitational-wave signals are detected in 15.5
days of live observation time; we set a frequentist upper limit
of 0.15 per day (at 90\% confidence level) on the rate of bursts
with large enough amplitudes to be detected reliably.
The amplitude sensitivity of the search, characterized using Monte
Carlo simulations, is several times better than that of previous
searches.
We also provide rough estimates of the distances at which
representative supernova and binary black hole merger signals could
be detected with 50\% efficiency by this analysis.
\end{abstract}

\pacs{
04.80.Nn, 
95.30.Sf, 
95.85.Sz  
}

\vspace{1.0ex} Submitted to {\it Classical and Quantum Gravity}

\section{Introduction}

Large interferometers are now being used to search for gravitational
waves with sufficient sensitivity to be able to detect signals from
distant astrophysical sources.  At present, the three detectors of the
Laser Interferometer Gravitational-wave Observatory (LIGO)
project~\cite{LIGOstatusAmaldi} have achieved
strain sensitivities consistent with
their design goals, while the GEO~600~\cite{GEOstatusAmaldi} and
Virgo~\cite{VIRGOstatusAmaldi} detectors are in the process of being
commissioned and are expected to reach comparable sensitivities.
Experience gained with these detectors,
TAMA300~\cite{TAMAstatusGWDAW04}, and several
small prototype interferometers has nurtured advanced designs for
future detector upgrades and new facilities, including Advanced
LIGO~\cite{AdLIGOFritschel}, Advanced Virgo~\cite{VIRGOstatusGWDAW05},
and the Large-scale Cryogenic
Gravitational-wave Telescope (LCGT) proposed to be constructed in
Japan~\cite{LCGTstatus03}.
The LIGO Scientific Collaboration (LSC) carries out the analysis of
data collected by the LIGO and GEO~600 gravitational-wave detectors,
and has begun to pursue joint searches with other collaborations
(see, for example, \cite{S2LIGOTAMA}) as
the network of operating detectors evolves.

As the exploration of the gravitational-wave sky can now be carried
out with greater sensitivity than ever before, it is important to
search for all plausible signals in the data.  In addition to
well-modeled signals such as those from binary
inspirals~\cite{Inspiral3PN} and spinning
neutron stars~\cite{JKS},
some astrophysical systems may emit gravitational waves which are
modeled imperfectly (if at all) and therefore cannot reliably be searched for
using matched filtering.  Examples of such imperfectly-modeled systems
include binary mergers (despite recent advances in the fidelity of
numerical relativity calculations for at least some
cases; see, for example,
\cite{Bakeretal})
and stellar core collapse events.  For the latter, several sets of simulations
have been carried out in the past (see, for example, \cite{DFM}
and~\cite{OBLW}), but more recent simulations have suggested a new
resonant core oscillation mechanism, driven by in-falling material,
which appears to power the supernova explosion and also to
emit strong gravitational waves~\cite{BLDOM,OBDL}.
Given the current uncertainties regarding gravitational wave emission
by systems such as these, as well as the possibility of detectable
signals from other astrophysical sources which are unknown or
for which no attempt has been
made to model gravitational wave emission, it is desirable to cast
a wide net.

In this article, we report the results of a search for gravitational-wave
``bursts'' that is designed to be able to detect short-duration
($\ll 1$~s) signals of arbitrary form as long as they have
significant signal power in the most sensitive frequency band of LIGO,
considered here to be $64$--$1600$~Hz.  This analysis uses LIGO data
from the fourth science run carried out by the LSC, called S4, and
uses the same basic methods as previous LSC burst
searches~\cite{s2burst,s3burst} that were performed using data from
the S2 and S3 science runs.  (A burst search was performed using data
from the S1 science run using different methods~\cite{s1burst}.)
We briefly describe the instruments and data collection in
section~\ref{sec:instruments}.
In sections~\ref{sec:triggers}
and~\ref{sec:consistency} we review the two complementary
signal processing methods---one based on locating signal power in
excess of the baseline noise and the other based on cross-correlating
data streams---that are used together to identify gravitational-wave
event candidates.
We note where the implementations have been improved relative to the
earlier searches and describe the signal consistency tests which are
based on the outputs from these tools.  Section~\ref{sec:select}
describes additional selection criteria which are used to ``clean up''
the data sample, reducing the average rate of spurious triggers in the data.
The complete analysis ``pipeline''
finds no event candidates that pass all of the selection criteria,
so we present in section~\ref{sec:results} an upper limit on the rate
of gravitational-wave events which would be detected reliably by our pipeline.

The detectability of a given type of burst, and thus the
effective rate limit for a particular astrophysical source
model, depends on the signal waveform and amplitude; in general, the
detection efficiency (averaged over sky positions and arrival times)
is less than unity.  We do not attempt a comprehensive survey of
possible astrophysical signals in this paper, but use a Monte Carlo
method with a limited number of ad-hoc
simulated signals to evaluate the amplitude sensitivity of our
pipeline, as described in section~\ref{sec:sensitivity}.  Overall,
this search has much better sensitivity than previous searches, mostly
due to using lower-noise data and partly due to improvements in the analysis
pipeline.
In section~\ref{sec:reach} we estimate the amplitude sensitivity for
certain modeled signals of interest and calculate approximate
distances at which those signals could be detected with 50\% efficiency.
This completed S4 search sets the stage for burst searches
now underway using data from the S5 science run of the LIGO and GEO~600
detectors, which benefit from much longer observation time and will
be able to detect even weaker signals.

\section{Instruments and data collection}  \label{sec:instruments}

\begin{figure}[bt]
\begin{center}
\includegraphics[width=10cm]{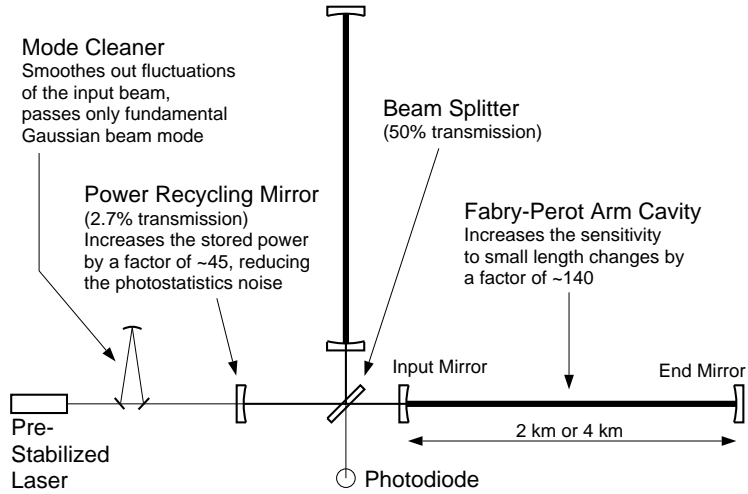}
\caption{Simplified optical layout of a LIGO interferometer.}
\label{fig:ifo}
\end{center}
\end{figure}

LIGO comprises two observatory sites in the United States with a total
of three interferometers.  As shown schematically in figure~\ref{fig:ifo},
the optical design is a Michelson interferometer augmented
with additional partially-transmitting mirrors to form Fabry-Perot
cavities in the arms and to ``recycle'' the outgoing beam power by
interfering it with the incoming beam.
Servo systems are used to ``lock'' the mirror positions to maintain
resonance in the optical cavities, as well as to control the mirror
orientations, laser frequency and intensity, and many other degrees of
freedom of the apparatus.
Interference between the two beams recombining at the beam splitter is
detected by photodiodes, providing a measure of the difference in arm
lengths that would be changed by a passing gravitational wave.
The large
mirrors which direct the laser beams are suspended from wires, with
the support structures isolated from ground vibrations using
stacks of inertial masses linked by damped
springs.  Active feed-forward and feedback systems provide additional suppression
of ground vibrations for many of the degrees of freedom.  The beam
path of the interferometer, excluding the laser light source and the
photodiodes, is entirely enclosed in a vacuum system.
The LIGO Hanford Observatory in Washington state has two
interferometers within the same vacuum system, one with arms 4~km long
(called H1) and the other with arms 2~km long (called H2).  The LIGO
Livingston Observatory in Louisiana has a single interferometer with
4~km long arms, called L1.

The response of an interferometer to a gravitational wave arriving at
local time $t$ depends on the dimensionless strain amplitude and
polarization of the wave and its arrival direction with respect to the
arms of the interferometer.
In the low-frequency limit, the differential strain
signal detected by the interferometer (effective arm length difference
divided by the length of an arm) can be expressed as a projection of
the two polarization components of the gravitational wave, $h_{+}(t)$
and $h_{\times}(t)$, with antenna response factors
$F_{+}(\alpha,\delta,t)$ and $F_{\times}(\alpha,\delta,t)$:
\begin{equation}\label{eq:hdet}
h_\mathrm{det}(t) = F_{+}(\alpha,\delta,t) \, h_{+}(t)
                  + F_{\times}(\alpha,\delta,t) \, h_{\times}(t) \, ,
\end{equation}
where $\alpha$ and $\delta$ are the right ascension and declination of
the source.  $F_{+}$ and $F_{\times}$ are distinct for each
interferometer site and change slowly with $t$ over the
course of a sidereal day as the Earth's rotation changes the
orientation of the interferometer with respect to the source location.

The electrical signal from the photodiode is filtered and digitized
continuously at a rate of 16\,384~Hz.  The time series of digitized
values, referred to as the ``gravitational-wave channel'' (GW channel),
is recorded
in a computer file, along with a timestamp derived from the Global
Positioning System (GPS) and additional information.  The relationship
between a given gravitational-wave signal and the digitized time
series is measured {\it in situ} by imposing continuous sinusoidal
position displacements of known amplitude on some of the mirrors.
These are called ``calibration lines'' because they appear as narrow
line features in a spectrogram of the GW channel.

Commissioning the LIGO interferometers has required several years of
effort and was the primary activity through late 2005.  Beginning in
2000, a series of short data collection runs was begun to establish
operating procedures, test the detector systems with stable
configurations, and provide data for the development of data analysis
techniques.  The first data collection run judged to have some
scientific interest, science run S1, was conducted in August-September
2002 with detector noise more than two orders of magnitude higher than
the design goal.  Science runs S2 and S3 followed in 2003 with
steadily improving detector noise, but with a poor duty cycle for L1
due primarily to low-frequency, large-amplitude ground motion from
human activities and weather.  During 2004, a hydraulic pre-isolation
system was installed and commissioned at the Livingston site to
measure the ground motion and counteract it with a relative
displacement between the external and internal support structures for
the optical components, keeping the internal components much closer to
an inertial frame at frequencies above $0.1$~Hz.  At the same time,
several improvements were made to the H1 interferometer at Hanford to
allow the laser power to be increased to the full design power of
$10$~W.

\begin{figure}[bt]
\begin{center}
\includegraphics[width=0.96\linewidth]{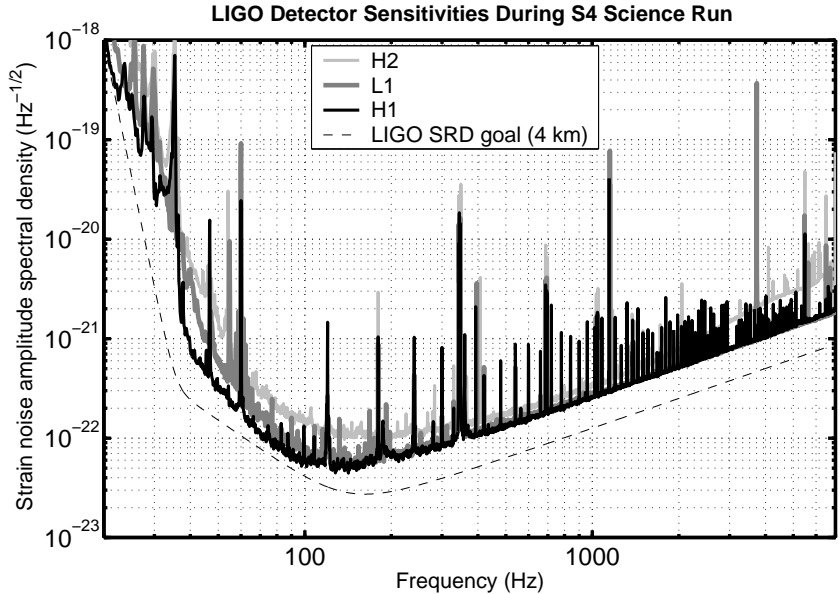}
\caption{Best achieved detector noise for the three LIGO
interferometers during the S4 science run, in terms of equivalent
gravitational wave strain amplitude spectral density.  ``LIGO SRD goal''
is the sensitivity goal for the 4-km LIGO interferometers set forth
in the 1995 LIGO Science Requirements Document~\cite{SRD}.}
\label{fig:s4noise}
\end{center}
\end{figure}

The S4 science run, which lasted from 22 February to 23 March 2005,
featured good overall ``science mode'' duty cycles of
$80.5$\%, $81.4$\%, and $74.5$\% for H1, H2, and L1, respectively, corresponding
to observation times of 570, 576, and 528~hours.
Thanks to the improvements made
after the S3 run, the detector noise during S4 was within a factor of
two of the design goal over most of the frequency band, as shown in
figure~\ref{fig:s4noise}.
The GEO~600 interferometer also collected data throughout the S4 run,
but was over a factor of 100 less sensitive than the LIGO interferometers
at $200$~Hz and a factor of few at and above the $1$~kHz frequency range.
The analysis approach used in this article effectively requires a
gravitational-wave signal to be distinguishable above the noise in
each of a fixed set of detectors, so it uses only the three LIGO
interferometers and not GEO~600.  There are a total of
402~hours
of S4 during which all three LIGO interferometers were simultaneously
collecting science-mode data.

\section{Trigger generation}  \label{sec:triggers}

The first stage of the burst search pipeline is to identify times when
the GW channels of the three interferometers appear to
contain signal power in excess of the baseline noise; these times,
along with parameters derived from the data, are called ``triggers''
and are used as input to later processing stages.  As in previous
searches~\cite{s2burst,s3burst}, the WaveBurst
algorithm~\cite{waveburst} is used for this purpose; it will only be
summarized here~\cite{WBCVS}.

WaveBurst performs a linear wavelet packet decomposition, using the symlet
wavelet basis~\cite{daubechies},
on short intervals of gravitational-wave data from each
interferometer.
This decomposition produces a time-frequency map of the data similar
to a windowed Fourier transformation.  A time-frequency data sample is
referred to as a pixel.
Pixels containing significant excess signal power are
selected in a non-parametric way by ranking them with other pixels at
nearby times and frequencies.
As in
the S3 analysis, WaveBurst has been configured for S4 to use six
different time resolutions and corresponding frequency resolutions,
ranging from $1/16$~s by $8$~Hz to $1/512$~s by $256$~Hz, to be able
to closely match the natural time-frequency properties of a variety of
burst signals.  The wavelet decomposition is restricted to
$64$--$2048$~Hz.  At any given resolution, significant pixels from the
three detector data streams are compared and coincident pixels are selected; these are
used to construct ``clusters'', potentially spanning many pixels in
time and/or frequency, within which there is evidence for a common
signal appearing in the different detector data streams.  These coincident clusters
form the basis
for triggers, each of which is characterized by a central time,
duration, central frequency, frequency range, and overall significance
$\Zg$ as defined in~\cite{WBperform}.
$\Zg$ is calculated from the pixels in the cluster and is roughly
proportional to the geometric average of the excess signal power
measured in the three interferometers, relative to the average
noise in each interferometer at the relevant frequency.  Thus,
a large value of
$\Zg$ indicates that the signal power in those pixels is highly
unlikely to have resulted from usual instrumental noise fluctuations.
In addition, the absolute strength of the signal detected by each
interferometer within the sensitive frequency band of the search
is estimated in terms of the root-sum-squared amplitude
of the detected strain,
\begin{equation}\label{eq:hdetrss}
\hrssdet =
     \sqrt{ \int \left| h_\mathrm{det}(t) \right|^2 \mathrm{d}t }  \, .
\end{equation}

WaveBurst was run on time intervals during which all three LIGO
interferometers were in science mode, but omitting periods when
simulated signals were injected into the interferometer hardware, any
photodiode readout experienced an overflow, or the data acquisition
system was not operating.  In addition, the last 30 seconds of each
science-mode data segment were omitted because it was observed that
loss of ``lock'' is sometimes preceded by a period of instability.
These selection criteria reduced the amount of data processed by WaveBurst from
402~hours to 391~hours.

\begin{figure}[bt]
\begin{center}
\includegraphics[width=0.60\linewidth]{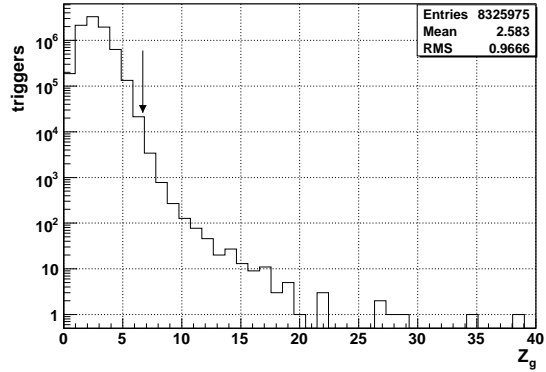}
\caption{Distribution of $\Zg$ values for all WaveBurst triggers.  The
arrow shows the location of the initial significance cut, $\Zg > 6.7$.}
\label{fig:Zghisto}
\end{center}
\end{figure}

For this analysis, triggers found by WaveBurst
are initially required to have a
frequency range which overlaps 64--1600~Hz.  An initial significance
cut, $\Zg \ge 6.7$, is applied to reject the bulk of the triggers and
limit the number passed
along to later stages of the analysis.
Figure~\ref{fig:Zghisto} shows the distribution of $\Zg$ prior to
applying this significance cut.

Besides identifying truly simultaneous signals in the three data
streams, WaveBurst applies the
same pixel matching and cluster coincidence tests to the three data
streams with many discrete relative time shifts imposed between the
Hanford and Livingston data streams, each much larger than the maximum
light travel time between the sites and the duration of the signals
targeted by this search.  The time-shifted triggers
found in this way provide a large sample to 
allow the ``background'' (spurious triggers produced in response to detector
noise in the absence of gravitational waves) to be studied,
under the assumption that the detector noise properties do not vary
much over the span of a few minutes and are independent at the two
sites.  The two Hanford data streams are {\em not} shifted relative to
one another, so that any local environmental effects which influence
both detectors are preserved.  In fact, some correlation in time is
observed between noise transients
in the H1 and H2 data streams.

\begin{figure}[bt]
\begin{center}
\includegraphics[width=0.96\linewidth]{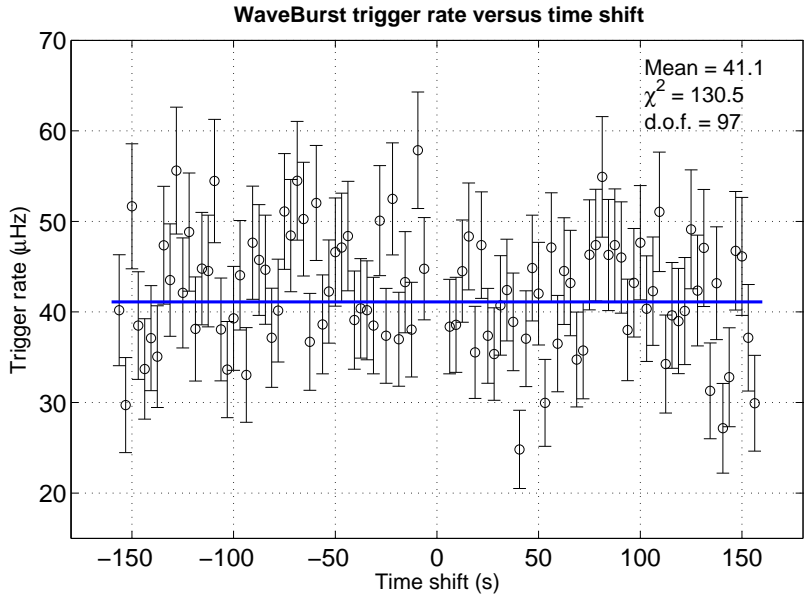}
\caption{WaveBurst trigger rate as a function of the relative time
shift applied between the Hanford and Livingston data streams.
The horizontal line is a fit to a constant value, yielding a $\chi^2$
of $130.5$ for 97 degrees of freedom.}
\label{fig:rateVSshift}
\end{center}
\end{figure}

Initially, WaveBurst found triggers for 98 time shifts in
multiples of $3.125$~s between $-156.25$ and $-6.25$~s and between
$+6.25$ and $+156.25$~s.  These 5119 triggers, called the ``tuning
set'', were used to choose the parameters of the signal consistency
tests and additional selection criteria described in the following two
sections.  As shown in figure~\ref{fig:rateVSshift},
the rate of triggers in the tuning set
is roughly constant for all time shifts, with a marginal $\chi^2$
value but without any gross dependence on time shift.
The unshifted triggers were kept hidden throughout
the tuning process, in order to avoid the possibility of human bias in
the choice of analysis parameters.

\section{Signal consistency tests}  \label{sec:consistency}

The WaveBurst algorithm requires only a rough consistency among the
different detector data streams---namely, some apparent excess power in the
same pixels in the wavelet decomposition---to generate a trigger.
This section describes more sophisticated consistency tests based on
the detailed content of the GW channels.  These tests
succeed in eliminating most WaveBurst triggers in the data, while
keeping essentially all triggers generated in response to
simulated gravitational-wave
signals added to the data streams.  (The simulation method is
described in section~\ref{sec:sensitivity}.)
Similar tests were also used in the S3 search~\cite{s3burst}.

\subsection{H1/H2 amplitude consistency test}  \label{subsec:ampcut}

\begin{figure}[bt]
\begin{center}
$
\begin{tabular}{cc}
\multicolumn{2}{c}{\includegraphics[width=0.49\linewidth]{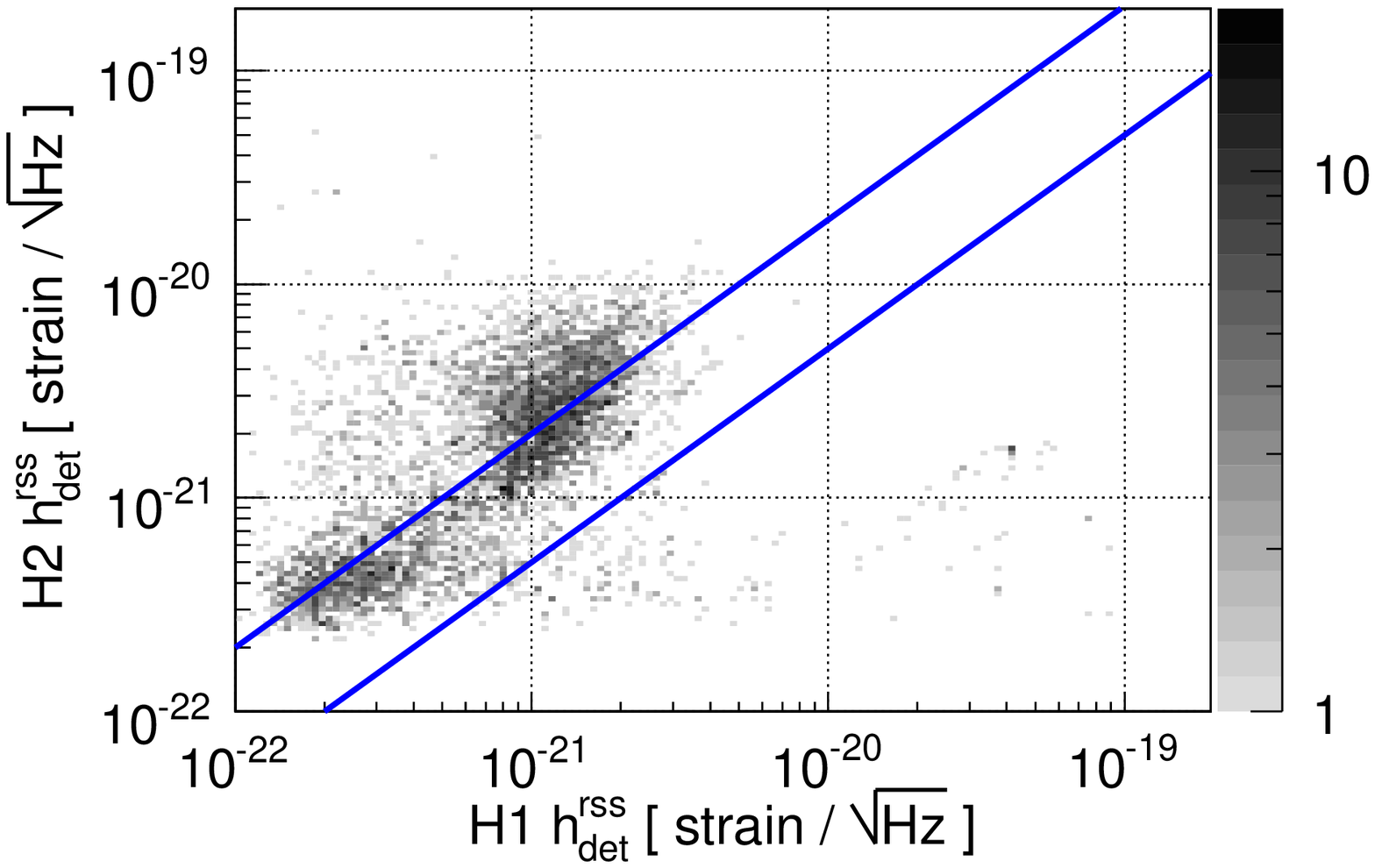} }\\
\multicolumn{2}{c}{\mbox{(a)} }\\

\includegraphics[width=0.465\linewidth]{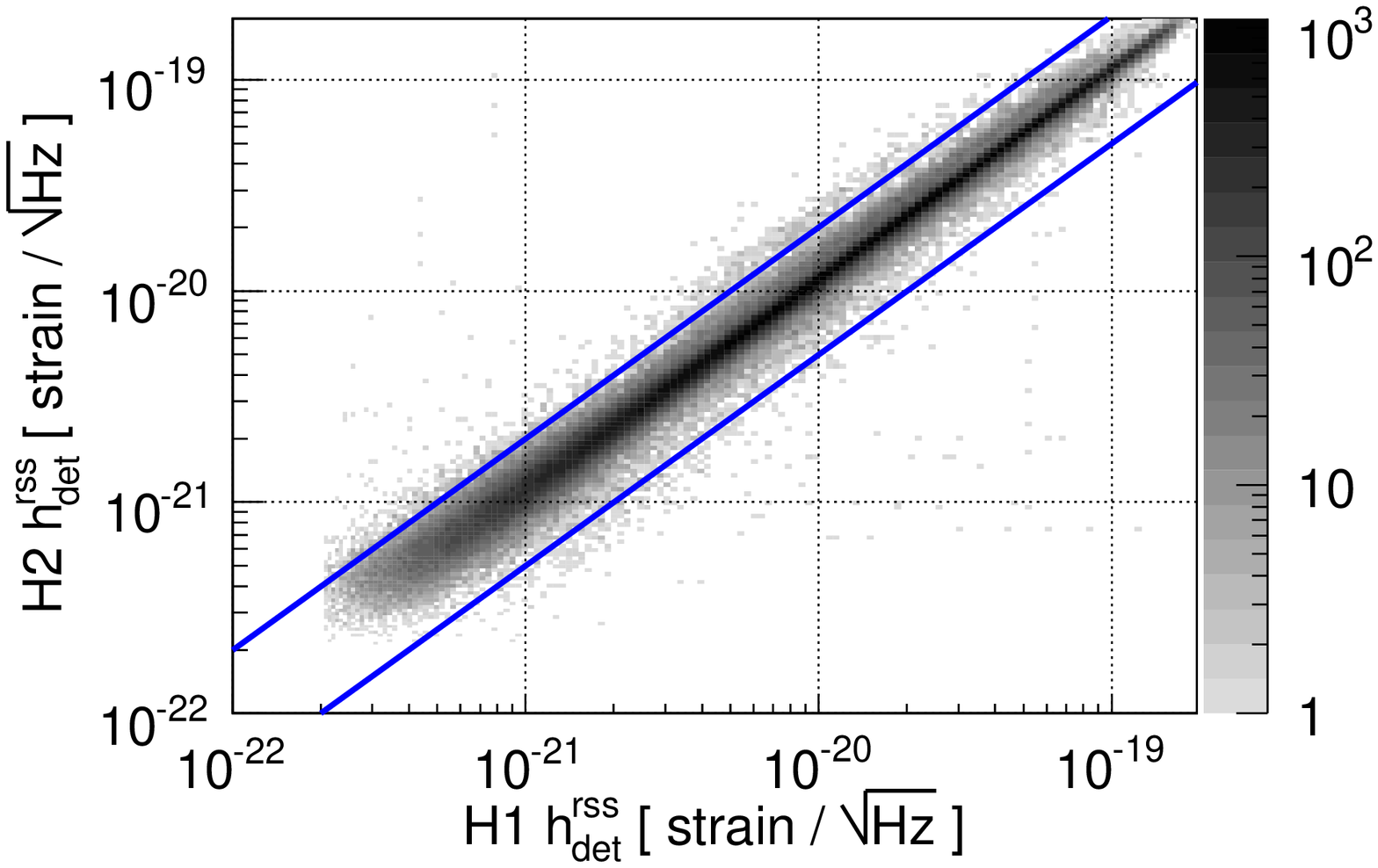} &
\includegraphics[width=0.465\linewidth]{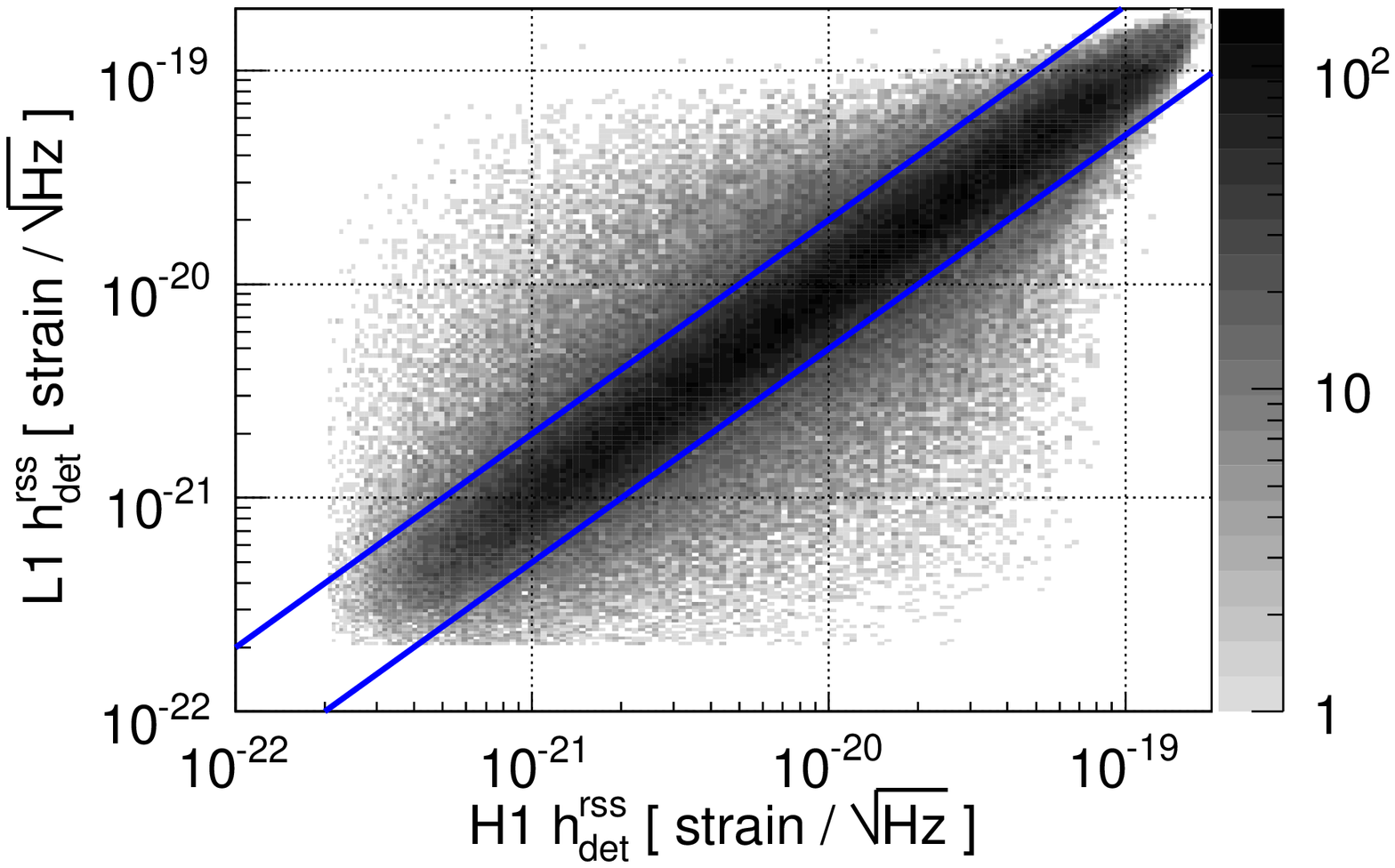} \\
\mbox{(b)} & \mbox{(c)} \\
\end{tabular}
$
\caption{(a) Two-dimensional histogram, with bin count indicated by greyscale,
of H2 vs.\ H1 amplitudes reconstructed
by WaveBurst for the tuning set of time-shifted triggers.
(b) Two-dimensional histogram of H2 vs.\ H1 amplitudes reconstructed for simulated
sine-Gaussian signals with a wide range of frequencies and
amplitudes from sources uniformly distributed over the sky
(see section~\ref{sec:sensitivity}).
In these plots, the diagonal lines show the limits of the H1/H2
amplitude consistency cut:
$0.5 < \mbox{ratio} < 2 \, .$
(c) Two-dimensional histogram of L1 vs.\ H1 amplitudes for the same simulated
sine-Gaussian signals.  Diagonal lines are drawn at ratios of $0.5$
and $2$ only to guide the eye; no cut is applied using this pair of
interferometers.}
\label{fig:H1H2amplTuningSG}
\end{center}
\end{figure}

Because the two Hanford interferometers are co-located and co-aligned, they will
respond identically (in terms of strain) to any given gravitational
wave.  Thus, the overall root-sum-squared amplitudes of the detected
signals, estimated
by WaveBurst according to equation (\ref{eq:hdetrss}), should agree well if
the estimation method is reliable.  Figure~\ref{fig:H1H2amplTuningSG}a
shows that the time-shifted triggers in the tuning set often have poor
agreement between the detected signal amplitudes in H1 and H2.  In
contrast, simulated signals injected into the data are
found with amplitudes which usually agree within a factor of 2, as
shown in figure~\ref{fig:H1H2amplTuningSG}b.  Therefore, we keep
a trigger only if the ratio of estimated signal amplitudes is in the
range $0.5$ to $2$.

The Livingston interferometer is roughly aligned with the Hanford
interferometers, but the curvature of the Earth makes exact alignment
impossible.  The antenna responses to a given gravitational wave will
tend to be similar, but not reliably enough to allow a consistency
test which is both effective at rejecting noise triggers and efficient
at retaining simulated signals, as shown in
figure~\ref{fig:H1H2amplTuningSG}c.

\subsection{Cross-correlation consistency tests}  \label{subsec:crosscorr}

The amplitude consistency test described in the previous subsection
simply compares scalar quantities derived from the data, without
testing whether the waveforms are similar in detail.  We use a program
called CorrPower~\cite{CorrPower}, also used in the S3 burst
search~\cite{s3burst}, to calculate statistics based on Pearson's
linear correlation statistic,
\begin{equation}
\label{eqn:rstat}
r = \frac {\sum_{i=1}^N (x_i - \bar{x})(y_{i} - \bar{y})} {\sqrt{\sum_{i=1}^N
    (x_i-\bar{x})^2}\sqrt{\sum_{i=1}^N (y_{i}-\bar{y})^2}} \, .
\end{equation}
In the above expression $\{ x_i \}$ and  $\{ y_i \}$
are sequences selected from the two GW channel time series,
possibly with a relative time shift,
and $\bar{x}$ and $\bar{y}$ are their respective mean values.
The length of each sequence, $N$ samples, corresponds to a chosen time
window (see below) over which the correlation is to be evaluated.
$r$ assumes values between $-1$ for fully anti-correlated
sequences and $+1$ for fully correlated sequences.

The $r$ statistic measures the correlation between two data streams,
such as would be produced by a common gravitational-wave signal
embedded in uncorrelated detector noise~\cite{rstat}.  It compares waveforms
without being sensitive to the relative amplitudes, and is thus
complementary to the H1/H2 amplitude consistency test described above.
Furthermore, the $r$ statistic may be used to test for a correlation
between H1 and L1 or between H2 and L1, even though these pairs
consist of interferometers with different antenna response factors,
because each polarization component will produce a
measurable correlation for a suitable relative time delay (unless
the wave happens to arrive from one of the special directions for
which one of the detectors has a null response for that polarization
component).  In the
special case of a linearly polarized gravitational wave, the detected
signals will simply differ by a multiplicative factor, which can be
either positive or negative depending on the polarization angle and
arrival direction.

Before calculating the $r$ statistic for each detector pair, the data
streams are filtered to select the frequency band of interest
(bandpass between 64~Hz and 1600~Hz) and whitened to equalize the
contribution of noise from all frequencies within this band.  The
filtering is the same as was used in the S3 search~\cite{s3burst}
except for the addition of a $Q$$=$$10$ notch filter, centered at
345~Hz, to avoid measuring correlations from the prominent
vibrational modes of the wires used to suspend the mirrors,
which are clustered around that frequency.
The $r$ statistic is then calculated over multiple time windows
with lengths of 20, 50, and 100~ms and a range of starting times,
densely placed (99\% overlap) to cover the full duration of the
trigger as reported by WaveBurst; the maximum value from among these
different time windows is used.

\hyphenation{lsc-group}
CorrPower~\cite{CPCVS}
calculates two quantities, derived from the $r$ statistic,
which are used to select triggers.  The first of these, called $R_0$,
is simply the signed cross-correlation between H1 and H2 with no
relative time delay.  Triggers with $R_0 < 0$ are rejected.  The second
quantity, called $\Gamma$, combines the $r$-statistic values from the
three detector pairs, allowing relative time delays of up to 11~ms
between H1 and L1 and between H2 and L1, and up to 1~ms between H1 and
H2 (to allow for a possible mismatch in time calibration).
Specifically, $\Gamma$ is the average of ``confidence'' values
calculated from the absolute value of each of the three individual
$r$-statistic values.
A large value of $\Gamma$ indicates that the data streams are
correlated to an extent that is highly unlikely to have resulted from
normal instrumental noise fluctuations.
This quantity complements $\Zg$, providing a different and largely
independent means for distinguishing real signals from background.

\begin{figure}[bt]
\begin{center}
$
\begin{tabular}{cc}
\includegraphics[width=0.465\linewidth]{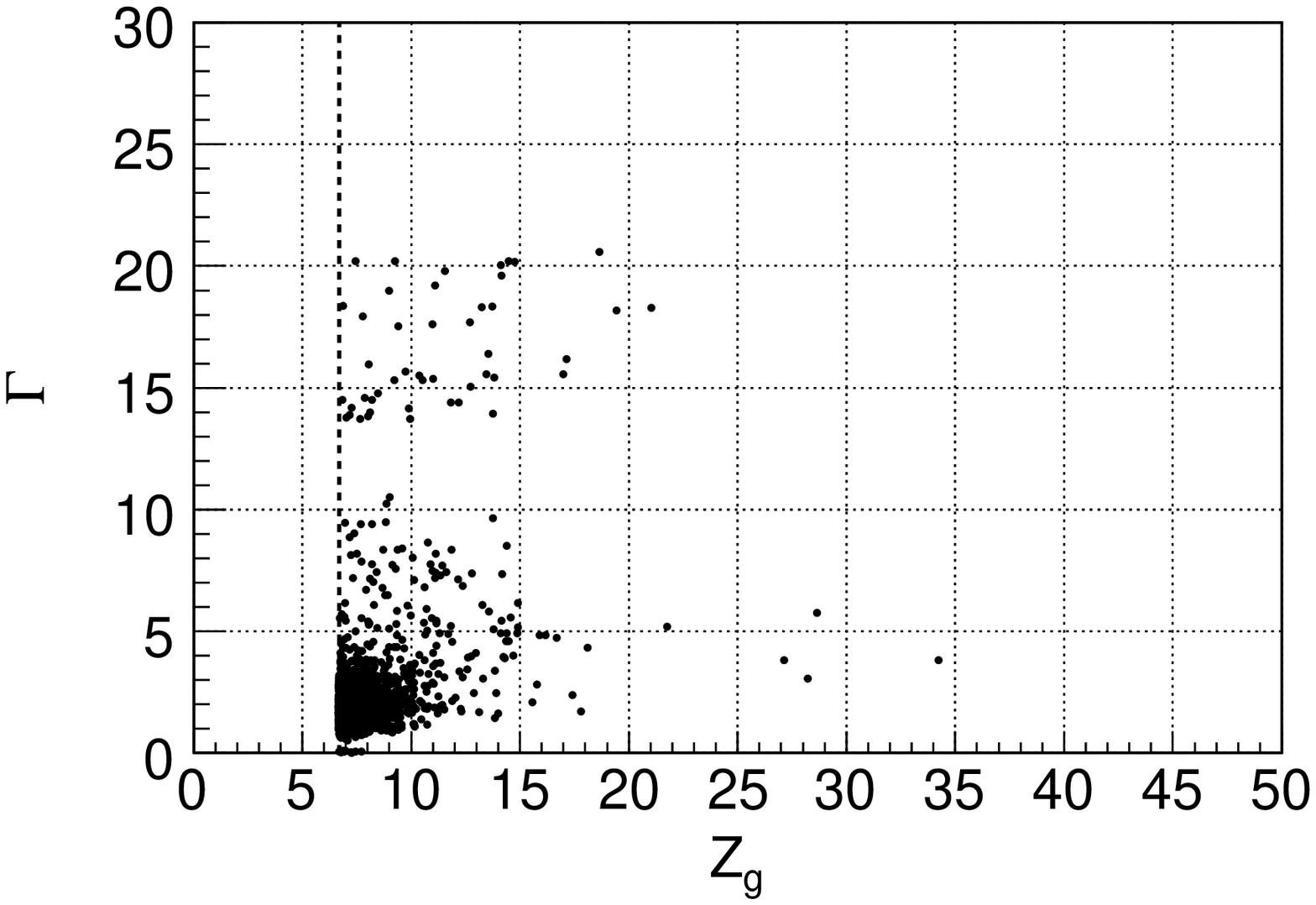} &
\includegraphics[width=0.465\linewidth]{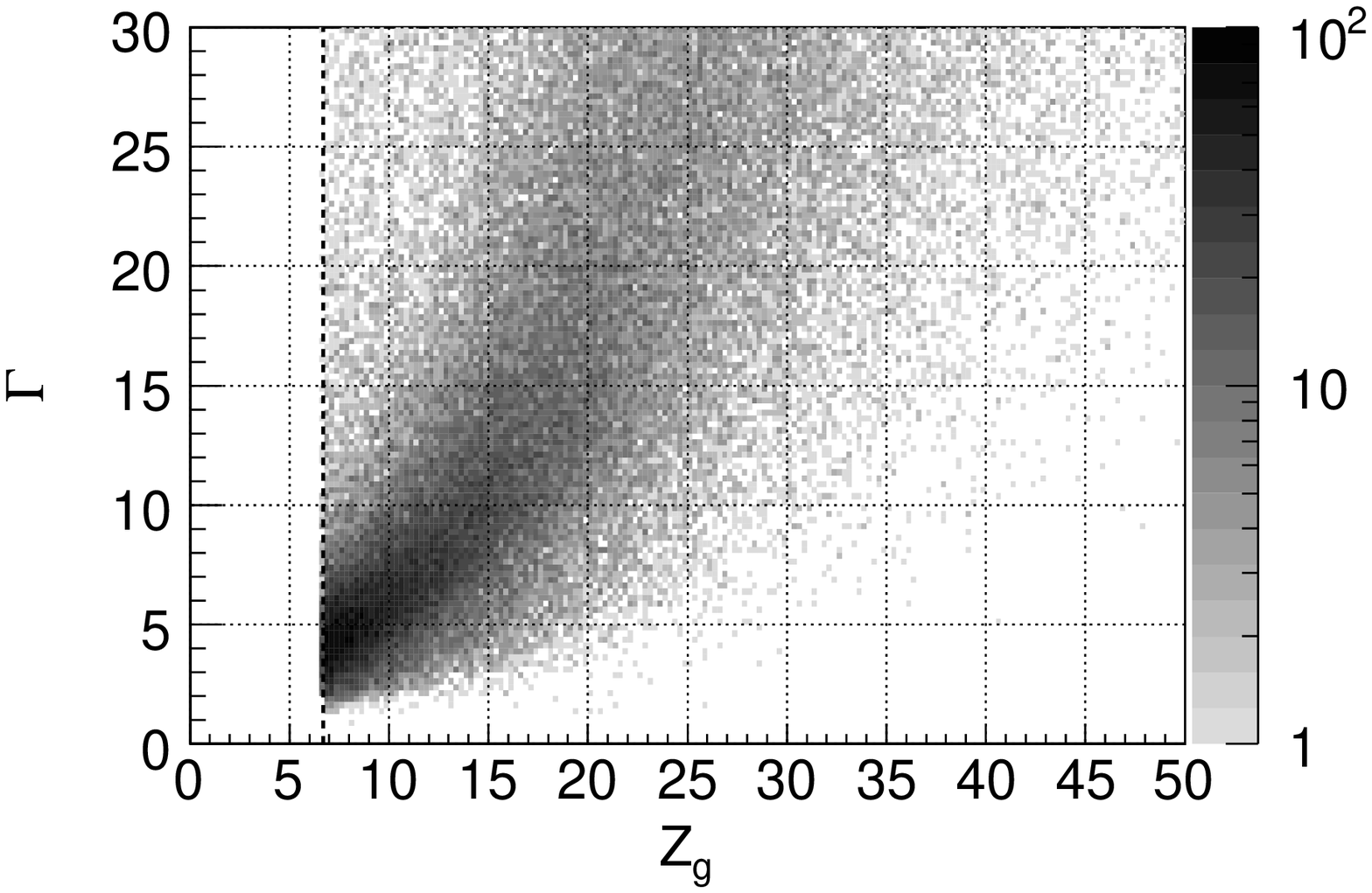} \\
\mbox{(a)} & \mbox{(b)} \\
\end{tabular}
$
\caption{Plots of $\Gamma$ versus $\Zg$, after the H1/H2
amplitude consistency cut but before any other cuts.
(a) Scatter plot for all time-shifted triggers in the tuning set.
(b) Two-dimensional histogram, with bin count indicated by greyscale,
for simulated sine-Gaussian
signals with a wide range of frequencies and
amplitudes from sources uniformly distributed over the sky (see
section~\ref{sec:sensitivity}).
In both plots, the vertical dashed line indicates the initial WaveBurst 
significance cut at $\Zg$$=$$6.7$.}
\label{fig:GammaZgTuningSG}
\end{center}
\end{figure}

Figure~\ref{fig:GammaZgTuningSG}
shows plots of $\Gamma$ vs.\ $\Zg$ for time-shifted triggers and for
simulated gravitational-wave signals after the H1/H2 amplitude
consistency cut but before the $R_0$ cut.
The time-shifted triggers with $\Gamma < 12$ and $\Zg < 20$ are the
tail of the
bulk distribution of triggers.  The outliers with $\Gamma > 12$ all
arise from a few distinct times when large noise transients occurred in H1 and
H2; these are found many times, paired with different L1 time shifts,
and have similar values of $\Gamma$ because the calculation of
$\Gamma$ is dominated by the H1-H2 pair in these cases.
The outliers with $\Gamma < 12$ and $\Zg > 20$ are artefacts of sudden
changes in the power line noise at $60$~Hz and $180$~Hz which
WaveBurst recorded as triggers.
A cut on the value of $\Gamma$
can eliminate many of the time-shifted triggers in
figure~\ref{fig:GammaZgTuningSG}a, but at the cost
of also rejecting weak genuine gravitational-wave signals that may
have the distribution in figure~\ref{fig:GammaZgTuningSG}b.
Therefore, the $\Gamma$ cut is
chosen only after additional selection criteria have been applied; see
section~\ref{subsec:gamma}.

\section{Additional selection criteria for event candidates}  \label{sec:select}

Environmental disturbances or instrumental misbehaviour occasionally
produce non-stationary noise in the GW channel of a
detector which contributes to the recording of a WaveBurst trigger.
These triggers can sometimes pass the H1-H2 consistency and
cross-correlation consistency tests, particularly since an
environmental disturbance at the Hanford site affects both H1 and H2.
As noted in the previous section, the calculated value of $\Gamma$ is
susceptible to being dominated by the H1-H2 pair even if there is
minimal signal power in the L1 data stream.  A significant background
rate of event candidates caused by environmental or instrumental
effects could obscure the rare gravitational-wave bursts that we seek,
or else require us to apply more aggressive cuts and thus lose
sensitivity for weak signals.

This section describes the two general tactics we use to reject data
with identifiable problems and thereby reduce the rate of background
triggers.  First, we make use of several ``data quality flags'' that
have been introduced in order to describe the status of the
instruments and the quality of the recorded data over time intervals
ranging from seconds to hours.  Second, we remove triggers attributed
to short-duration instrumental or environmental effects by applying
``vetoes'' based on triggers generated from auxiliary channels which have
been found to correlate with transients in the GW channel.
Applying data quality conditions and vetoes to the data set reduces
the amount of ``live'' observation time (or ``livetime'') during which
an arriving gravitational-wave burst would be detected and kept as an
event candidate at the end of the analysis pipeline.  Therefore, we
must balance this loss (``deadtime'') against the effectiveness for
removing spurious triggers from the data sample.

Choosing data quality and veto conditions with reference to a sample
of gravitational-wave event candidates could introduce a selection
bias and invalidate any upper limit calculated from the sample.
Therefore, we have evaluated the relevance of potential data quality cuts and
veto conditions using other trigger samples.  In addition to
the tuning set of time-shifted WaveBurst triggers, we have applied the
KleineWelle~\cite{KWQLPEF} method to identify transients in each
interferometer's GW channel.
(We have also used KleineWelle to identify transients in numerous
auxiliary channels for veto studies, as described in~\ref{subsec:vetoes}.)
Like WaveBurst,
KleineWelle is a time-frequency method utilizing multi-resolution
wavelet decomposition,
but it processes each data channel independently~\cite{KWCVS}.
In analyzing data, the time series is first whitened using a linear 
predictor error filter~\cite{KWQLPEF}.
Then the time-frequency decomposition is obtained using the Haar
wavelet transform.
The squared wavelet coefficients normalized to 
the scale's (frequency's) root-mean-square provide an estimate of the
energy associated with a certain time-frequency pixel.
A clustering mechanism is invoked in order to increase the sensitivity
to signals with less than optimal shapes in the time-frequency plane
and a total normalized cluster energy is computed.
The significance of a cluster is then defined as the negative natural
logarithm of the probability of the computed total normalized cluster
energy to have resulted from Gaussian white noise; we apply
a threshold on this significance to define KleineWelle triggers.
The samples of KleineWelle triggers from each detector, as well as
the subsample of coincident H1 and H2 triggers, are useful indicators
of localized disturbances.  They may in principle contain one or more
genuine gravitational-wave signals, but decisions about data quality
and veto conditions are based on the statistics of the entire sample
which is dominated by instrumental artefacts and noise fluctuations.

\subsection{Data quality conditions}  \label{subsec:dq}

We wish to reject
instances of clear hardware problems with the LIGO detectors
or conditions that could affect our ability to unequivocally register
the passage of gravitational-wave bursts.
Various studies of the data, performed during and after data
collection, produced a catalog of conditions that might affect the
quality of the data.  Each named condition, or ``flag'', has an
associated list of time intervals during which the condition is
present, derived either from one or more diagnostic channels or from
entries made in the electronic logbook by operators and scientific
monitors.
We have looked for significant correlations
between the flagged time intervals and time-shifted WaveBurst triggers,
and also between the flagged time intervals and KleineWelle
single-detector triggers (particularly the ``outliers'' with large
significance and the coincident H1 and H2 triggers).
Based on these studies, we decided to impose a number of
data quality conditions.

We first require the calibration lines to be continuously present.
On several occasions when they dropped out briefly, due to a
problem with the excitation engine, the data is
removed from the analysis.
The livetime associated with these occurrences is negligible
while they are all correlated with transients appearing
in the GW channel.

Local winds and sound from airplanes
may couple to the instrument through
the ground and result in elevated noise and/or impulsive signals.
A data quality flag was established to
identify intervals of local winds at the sites with speeds of
$56$~km/hour (35 miles per hour) and above.
We studied the correlation of these times with
the single-detector triggers produced with KleineWelle.
The correlation is more apparent in the H2 detector, for which 7.4\% of
the most significant KleineWelle triggers (threshold of 1600) 
coincide with the intervals of
strong winds at the Hanford site.
The livetime that is rejected in this way is 0.66\% of the H1-H2 coincident
observation time over which this study was performed.
Thanks to improved acoustic isolation installed after the S2 science
run, acoustic noise from airplanes was not found to contribute to
triggers in the GW channel in general; however,
a period of 300 seconds has been rejected around a particularly loud
time when a fighter jet passed over the Hanford site.

Elevated low-frequency seismic activity has been observed to cause
noise fluctuations and transients in the GW channel.
Data from several seismometers at the Hanford observatory was
band-pass filtered in various narrow bands between $0.4$~Hz and
$2.4$~Hz, and the root-mean-square signal in each band was tracked
over time.  A set of particularly relevant seismometers and bands was
selected, and time intervals were flagged whenever a band in this set
exceeded 7 times its median value.
A follow up analysis of the single instrument as well as coincident
H1-H2 KleineWelle triggers
found significant correlation with the elevated seismic noise.
The strongest correlation is observed in the outlier triggers
(KleineWelle 
significance of 1600 or greater) in H2, of which 41.9\% coincide with the
seismic flags, compared to a deadtime of 0.6\%.

In the two Hanford detectors, a diagnostic channel counting ADC overflows in the
length sensing and control subsystem was used to flag 
intervals for exclusion from the analysis.
One minute of livetime around these overflows is rejected.
Such overflows were indeed seen to correlate with single-detector
outlier triggers in H1 (44.4\% of them, with 0.68\% deadtime) and H2
(74.1\% of them, with 0.41\% deadtime).

Two data quality cuts are derived from ``trend'' data (summaries of
minimum, maximum, mean and root-mean-square values over each
one-second period) monitoring the interferometry
used in the LIGO detectors.
The first one is based on occasional transient dips in the stored
light in the arm cavities.
These have been identified by scanning the trend data for the relevant
monitoring photodiodes, defining the size of a dip as the fractional drop of
the minimum in that second relative to the average of the previous
ten seconds, and applying various thresholds on the minimum dip size.
For the three LIGO detectors, thresholds of 5\%, 4\%
and 5\% respectively for L1, H1 and H2 are used.
High correlation of such light dips with single-detector triggers
is observed, while
the deadtime resulting from them in each of the three LIGO instruments
is less than 0.6\%.
The second data quality cut of this type is based on the DC level
of light reaching the photodiode at the output of the interferometer,
which sees very little light when the interferometer is operating properly.
By thresholding on the trend data for this channel, intervals when
its value was unusually high are identified in H1 and L1.
These intervals are seen to correlate with instrument outlier triggers
significantly. The deadtime resulting from them is 1.02\% in H1 and
1.74\% in L1.

Altogether, these data quality cuts result in a net loss of observation time
of 5.6\%.

\subsection{Auxiliary-channel vetoes}  \label{subsec:vetoes}

LIGO records thousands of auxiliary read-back channels of the
servo control systems employed in the instruments' interferometric
operation as well as auxiliary channels monitoring the instruments'
physical environment.
There are plausible couplings of environmental disturbances or servo
instabilities both to these monitoring channels and to the GW channel;
thus, transients appearing in these auxiliary channels may be used to veto
triggers seen simultaneously in the GW channel.
This assumes that a genuine gravitational-wave burst would not appear
in these auxiliary channels, or at least that any coupling is small
enough to stay below the threshold for selecting transients in these
channels.

We have used KleineWelle to produce triggers from
over 100 different auxiliary channels that monitor the interferometry
and the environment in the three LIGO detectors.  
A first analysis of single-detector KleineWelle triggers from the L1
GW channel and coincident KleineWelle triggers
from the H1 and H2 GW channels against 
respective auxiliary channels identified the ones that showed high
GW channel trigger rejection power with minimal livetime loss
(in the vast majority of channels much less that 1\%).
In addition to interferometric channels, environmental ones
(accelerometers and microphones) located on the optical tables holding the
output optics and photodiodes
appeared to correlate with GW channel triggers
recorded at the same site.

Auxiliary interferometric channels (besides the GW
channel) could in principle be affected by a gravitational wave,
and a veto condition derived from such a channel could reject a
genuine signal.
Hardware signal injections
imitating the passage of gravitational waves
through our detectors, performed at several pre-determined times
during the run, have been used to establish
under what conditions each channel is safe to use as a veto.
Non-detection of a hardware injection
by an auxiliary channel 
suggests the unconditional safety of this channel as a veto in the search,
assuming that a reasonably
broad selection of signal strengths and frequencies were injected.
But even if hardware injections are seen in the auxiliary channels,
conditions  can readily be
derived under which no triggers caused by the hardware injections
are used as vetoes.
This involves imposing conditions on the significance of the trigger
and/or on the ratio of the signal strength seen in
the auxiliary channel to that seen in the GW channel.
We have thus established the conditions under which 
several channels involved in the
length and angular sensing and control systems of the interferometers
can be used safely as vetoes.
(The data quality conditions described in section~\ref{subsec:dq} were
also verified to be safe using hardware injections.)

The final choice of vetoes was made by examining the tuning set of
time-shifted triggers remaining in the WaveBurst search pipeline
after applying the signal consistency tests and data quality conditions.
The ten triggers from the time-shifted analysis with the largest
values of $\Gamma$, plus the ten with the largest values of $\Zg$,
were examined and six of them
were found to coincide with transients in one or more of
the following channels:
the in-phase and quadrature-phase demodulated signals
from the pick-off beam from the H1 beamsplitter,
the in-phase demodulated pitch signal from one of
the wavefront sensors used in the H1 alignment sensing and control system,
the beam splitter pitch and yaw control signals,
and accelerometer readings
on the optical tables holding the H1 and H2
output optics and photodiodes.
KleineWelle triggers produced from these seven auxiliary channels were
clustered (with a 250~ms window) and their union was taken.
This defines the final list of veto triggers for this search,
each indicating a time interval (generally $\ll 1$~s long) to be vetoed.

The total duration of the veto triggers considered in this analysis is
at the level of 0.15\% of the total livetime. However, this does not
reliably reflect the deadtime of the search since a GW
channel trigger is vetoed if it has any overlap with a veto trigger.
Thus, the actual deadtime of the search depends on the duration of the
signal being sought, as reconstructed by WaveBurst.
We reproduce this effect in the Monte Carlo simulation used to
estimate the efficiency of the search (described in
section~\ref{sec:sensitivity}) by applying the same analysis pipeline
and veto logic.
The effective deadtime depends on the morphology of the signal and on
the signal amplitude, since larger-amplitude signals tend to be
assigned longer durations by WaveBurst.
For the majority of waveforms we considered in this search and for 
plausible signals strengths, the resulting effective deadtime is of the
order of 2\%.
Because this loss is signal-dependent, in this analysis we consider it
to be a loss of efficiency rather than a loss of live observation
time; in other words, the live observation time we state reflects the
data quality cuts applied but does not reflect the auxiliary-channel
vetoes.

\subsection{Gamma cut}  \label{subsec:gamma}

\begin{figure}[bt]
\begin{center}
$
\begin{tabular}{cc}
\includegraphics[width=0.465\linewidth]{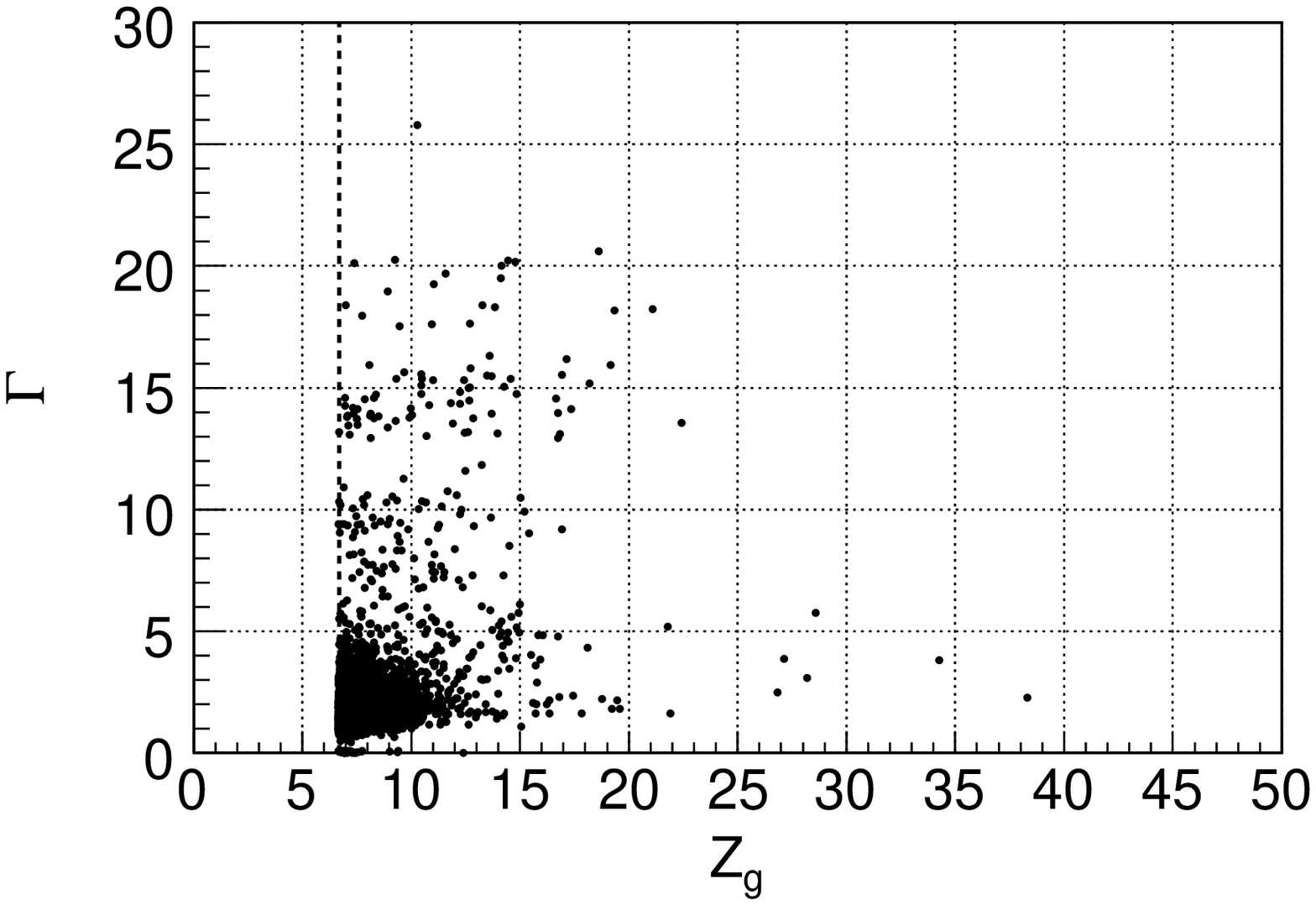} &
\includegraphics[width=0.465\linewidth]{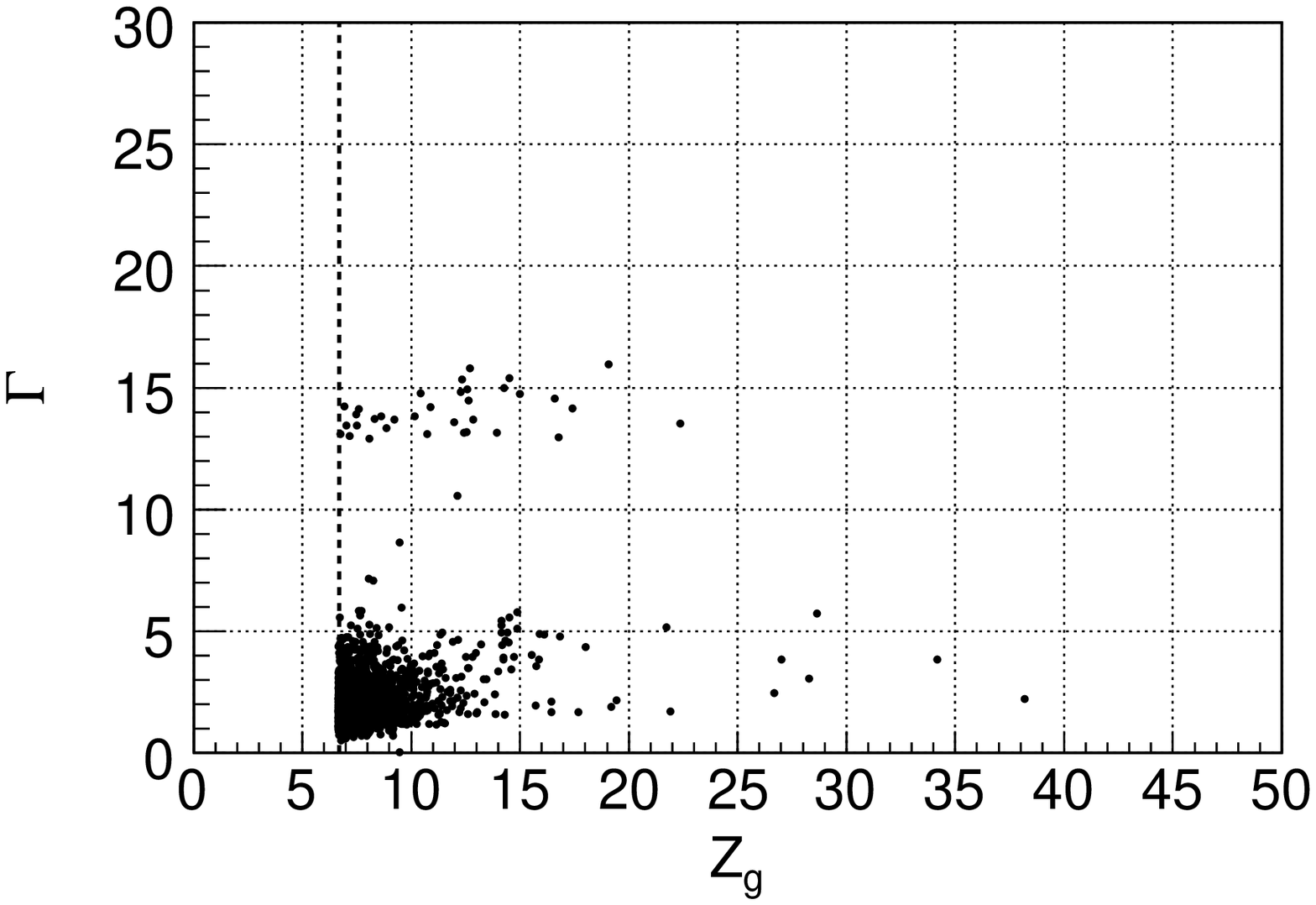} \\
\mbox{(a)} & \mbox{(b)} \\
\includegraphics[width=0.465\linewidth]{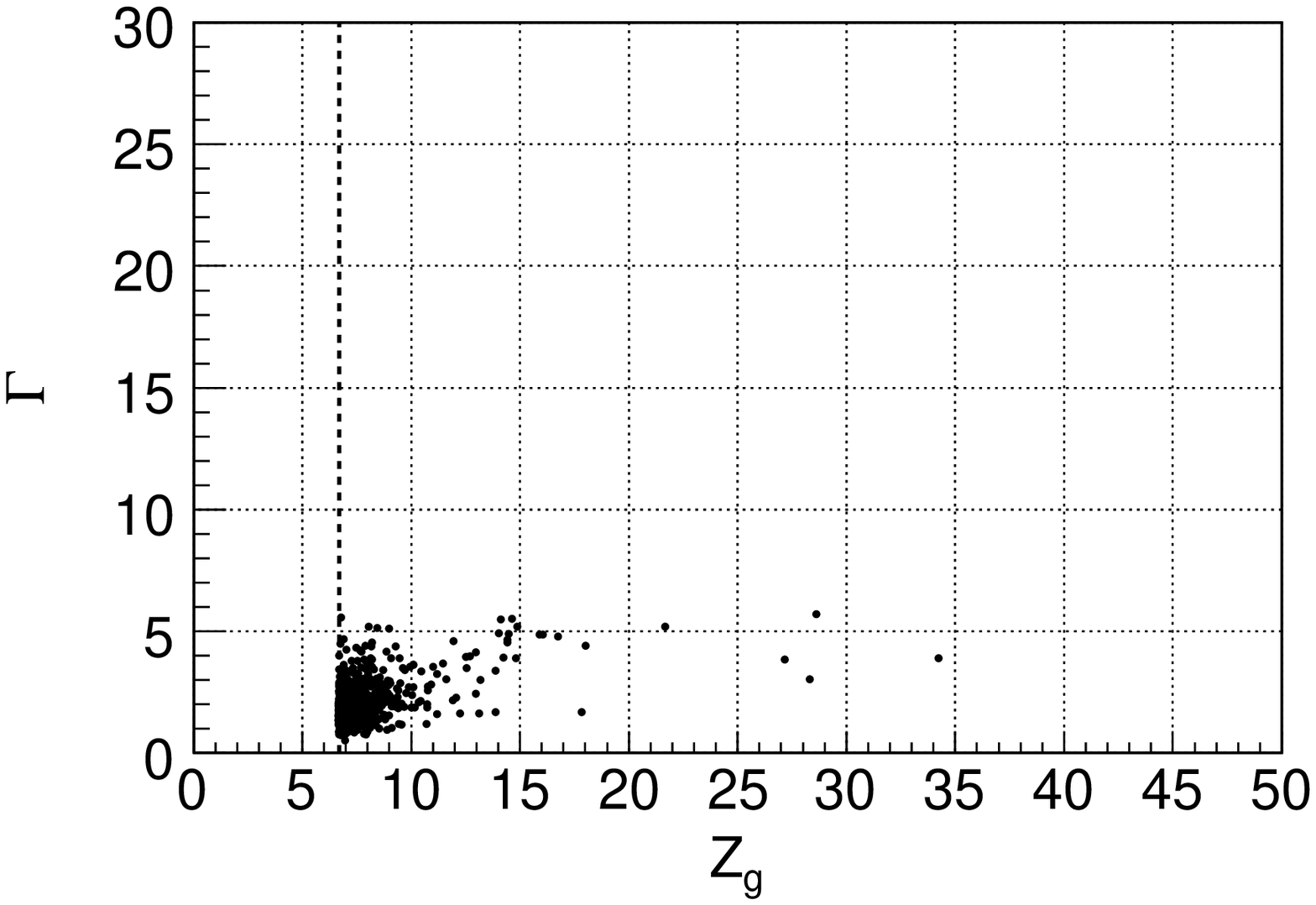} &
\includegraphics[width=0.465\linewidth]{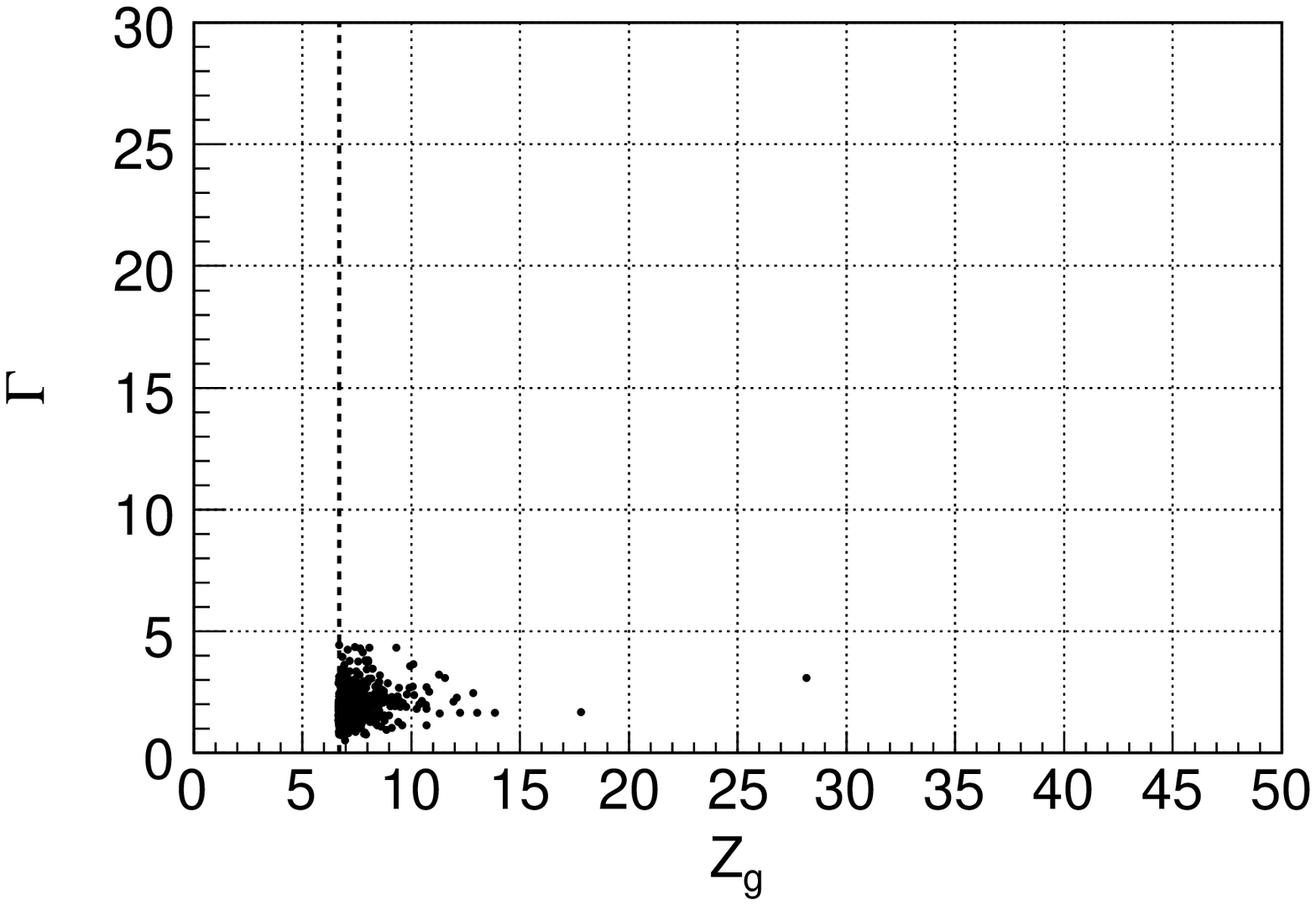} \\
\mbox{(c)} & 
\mbox{(d)} \\
\end{tabular}
$
\caption{Scatter plots of $\Gamma$ versus $\Zg$ for the tuning set of
time-shifted triggers. (a) All triggers; (b) after data quality cuts;
(c) after data quality and H1-H2 consistency cuts (amplitude ratio and
$R_0$); (d) after data quality, H1-H2 consistency, and auxiliary-channel
vetoes.}
\label{fig:GammaZgTuning}
\end{center}
\end{figure}

The cuts described above cleaned up the outliers in the data
considerably, as shown by the sequence of scatter plots in
figure~\ref{fig:GammaZgTuning}.
Following the data quality and veto criteria we just described, the
remaining time-shifted WaveBurst triggers (shown in
figure~\ref{fig:GammaZgTuning}d)
were used as the basis for choosing the cross
correlation $\Gamma$ threshold.
As with previous all-sky searches for gravitational-wave bursts
with LIGO, we desire the number of background triggers expected
for the duration of the observation to be much less than 1 but not
zero, typically of order $\sim 0.1$.
On that basis, we chose a threshold of $\Gamma >4$ which results
in 7 triggers in 98 time shifts,
or 0.08 such triggers normalized to the duration of the S4 observation
time.

\section{Search results}  \label{sec:results}

\begin{figure}[bt]
\begin{center}
$
\begin{tabular}{cc}
\includegraphics[width=0.465\linewidth]{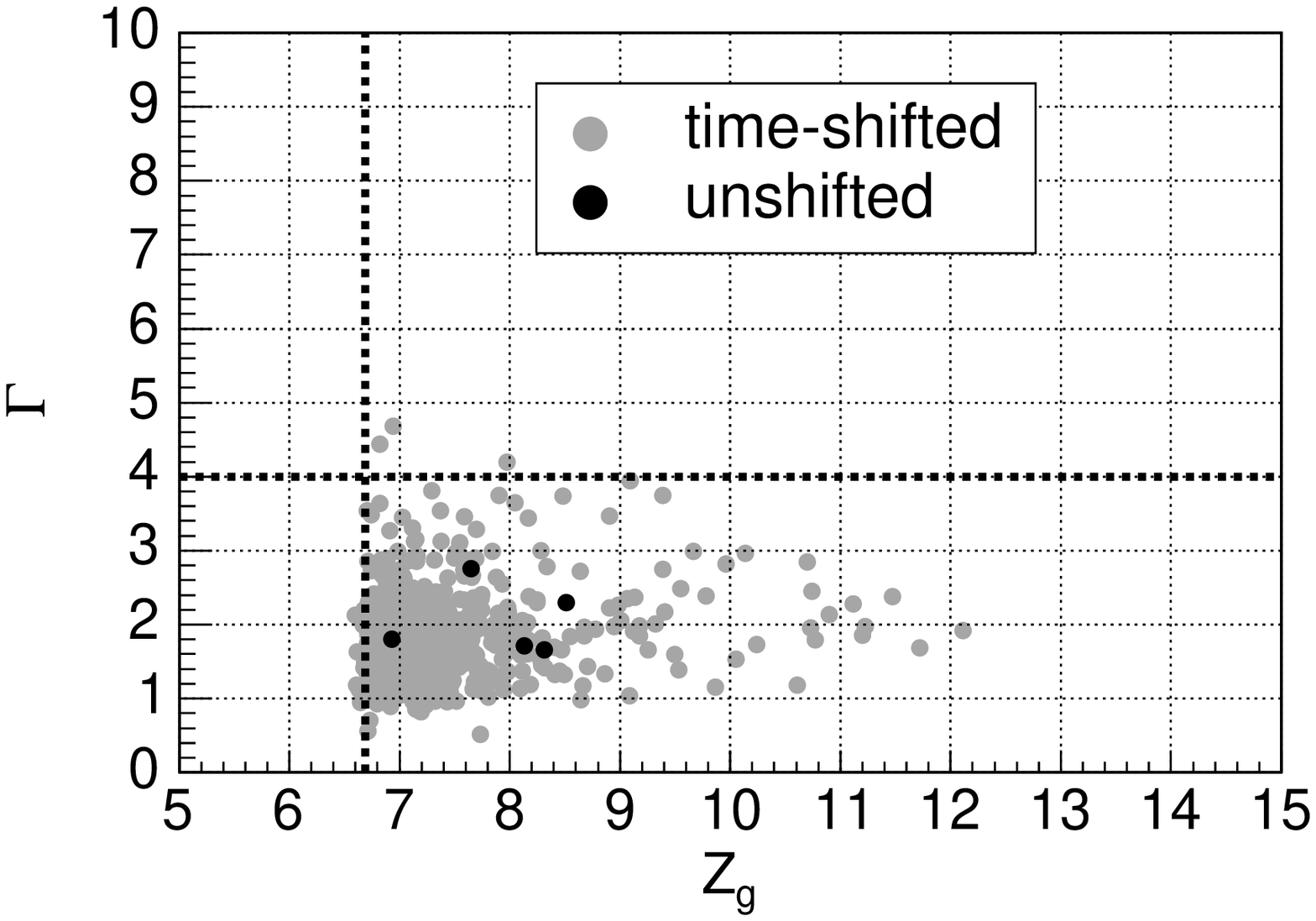} &
\includegraphics[width=0.465\linewidth]{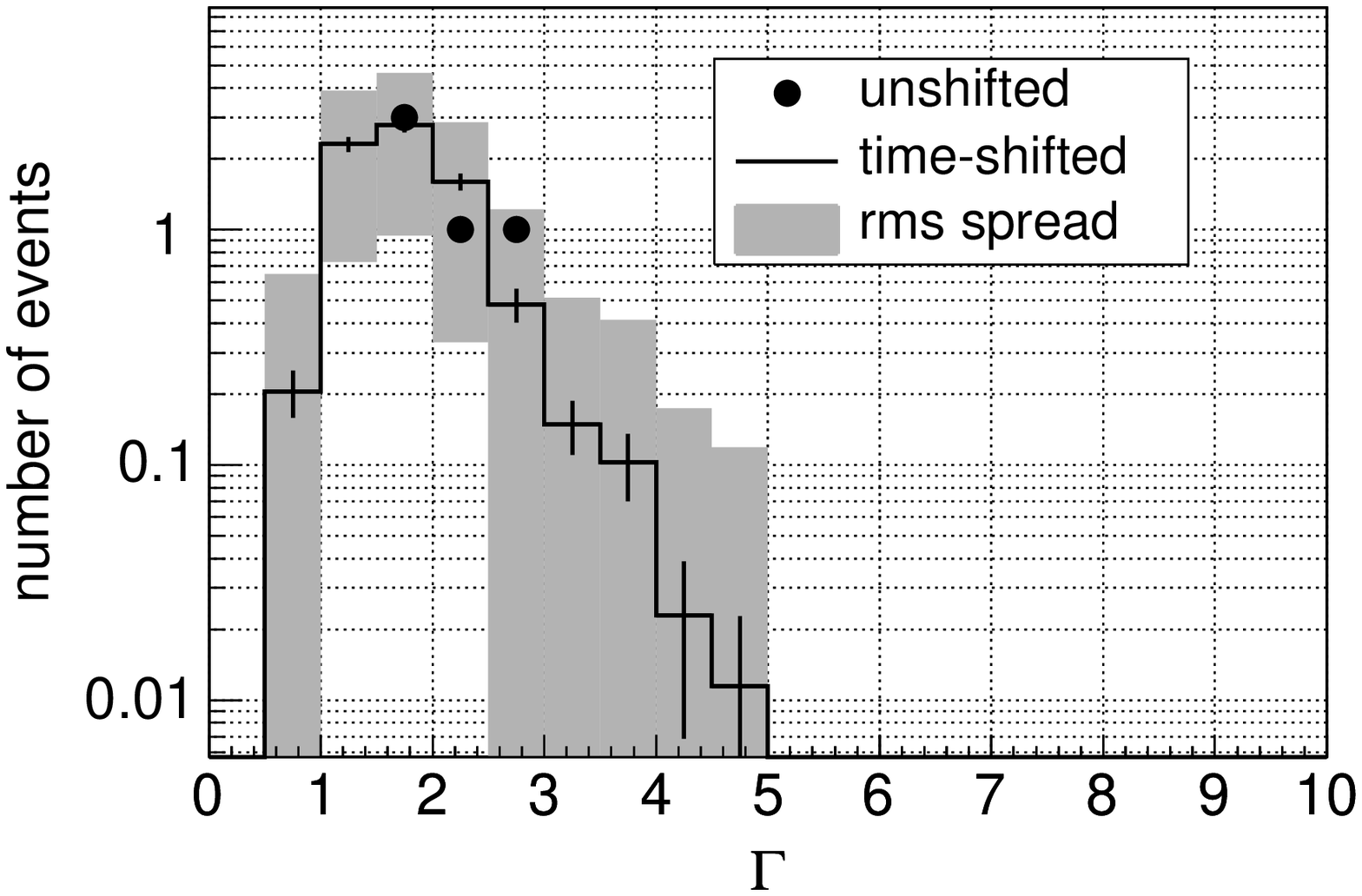} \\
\mbox{(a)} & \mbox{(b)} \\
\end{tabular}
$
\caption{(a) Scatter plot of $\Gamma$ vs.\ $\Zg$ for time-shifted
triggers (grey circles) and unshifted triggers (black circles)
after all other analysis cuts.  The vertical dashed line indicates the
initial WaveBurst significance cut at $\Zg$$=$$6.7$.
The horizontal dashed line indicates the final $\Gamma$ cut.
(b) Overlaid
histograms of $\Gamma$ for unshifted triggers (black circles) and mean
background estimated from time-shifted triggers (black stairstep
with statistical error bars).  The shaded bars represent the
expected root-mean-square statistical fluctuations on the number of
unshifted background triggers in each bin.}
\label{fig:GammaZgResult}
\end{center}
\end{figure}

After all of the trigger selection criteria had been established using
the tuning set of time-shifted triggers, WaveBurst was
re-run with a new, essentially independent set of 100 time shifts, in
increments of 5~s from $-250$~s to $-5$~s and from $+5$~s to $+250$~s,
in order to provide an estimate of the background which is minimally biased
by the choice of selection criteria.  The total effective livetime
for the time-shifted sample is $77.4$ times the unshifted observation
time, reflecting the reduced overlap of Hanford and Livingston data
segments when shifted relative to one another.
The unshifted triggers 
were looked at for the first time.
Table~\ref{t:cutcounts}
\begin{table}
\caption[]{Counts of time-shifted and unshifted triggers as cuts are
applied sequentially.  The column labeled ``Normalized'' is the
time-shifted count divided by 77.4, representing an estimate of the
expected background for the S4 observation time.}
\label{t:cutcounts}
\begin{indented}
\item[]\begin{tabular}{lccc}
\br
 & \centre{2}{Time-shifted} \\ \ns
 & \crule{2} & Unshifted \\
Cut & Count & Normalized & Count \\
\mr
Data quality & 3153 & $40.7$ & 44 \\
H1/H2 amplitude consistency & 1504 & $19.4$ & 14 \\
$R_0 > 0$ & ~755 & $~9.8$ & ~5 \\
Auxiliary-channel vetoes & ~671 & $~8.7$ & ~5 \\
$\Gamma > 4$ & ~~~3 & $0.04$ & ~0 \\
\br
\end{tabular}
\end{indented}
\end{table}
summarizes the trigger counts for these time-shifted and unshifted
triggers at each stage in the sequence of cuts.  In addition, the
expected background at each stage (time-shifted triggers normalized to
the S4 observation time) is shown for direct comparison with the
observed zero-lag counts.
Figure~\ref{fig:GammaZgResult}
shows a scatter plot of $\Gamma$ vs.\ $\Zg$ and histograms of
$\Gamma$ for both time-shifted and unshifted triggers after all other
cuts.
These new time-shifted triggers are statistically consistent with the
tuning set (figure~\ref{fig:GammaZgTuning}d), although no triggers
are found with $\Zg > 15$ in this case.
Five unshifted triggers are found, distributed in a manner reasonably
consistent with the background.  All five have $\Gamma$$<$$4$ and thus
fail the $\Gamma$ cut.  Three time-shifted
triggers pass the $\Gamma$ cut, corresponding to an
estimated average background of $0.04$ triggers over the S4 observation
time.

With no unshifted triggers in the final sample, we place an upper
limit on the mean rate of gravitational-wave events that
would be detected reliably ({\it i.e.}, with efficiency near unity) by
this analysis pipeline.  Since the background estimate is small and is
subject to some systematic uncertainties, we simply take it to be zero
for purposes of calculating the rate limit; this makes the rate limit
conservative.  With 15.5 days of
observation time, the one-sided frequentist upper limit on the rate at
90\% confidence level is
$-\ln{(0.1)}/T$ = $2.303/(15.5~\mbox{days}) = 0.15$ per day.
For comparison, the S2 search~\cite{s2burst}
arrived at an upper limit of $0.26$ per day.
The S3 search~\cite{s3burst} had an observation time of only
$8$~days and did not state a rate limit.

\section{Amplitude sensitivity of the search}  \label{sec:sensitivity}

The previous section presented a limit on the rate of a hypothetical
population of gravitational-wave signals for which the analysis
pipeline has perfect detection efficiency.  However, the actual
detection efficiency will depend on the signal waveform and amplitude,
being zero for very weak signals and generally approaching unity for
sufficiently strong signals.
The signal processing methods used in this analysis are expressly
designed to be able to detect arbitrary waveforms as long as they have
short duration and frequency content in the $64$--$1600$~Hz band which
stands out above the detector noise.  Therefore, for any given signal
of this general type, we wish to determine a characteristic minimum
signal amplitude for which the pipeline has good detection efficiency.
As in past analyses, we use a Monte Carlo technique with a
population of simulated gravitational wave sources.  Simulated events
are generated at random sky positions and pseudo-random times
(imposing a minimum separation of 80~s) during the S4 run; the resulting
signal waveforms in each interferometer are calculated with the
appropriate antenna factors and time delays.  These simulated signals
are added to the actual detector data, and the summed data streams are
analyzed using the same pipeline with the same trigger selection
criteria.

The intrinsic amplitude of a simulated gravitational wave may be
characterized by its root-sum-squared strain amplitude {\em at the
Earth}, without folding in antenna response factors:
\begin{equation}\label{eq:hrss}
\hrss \equiv \sqrt{\int (|h_{+}(t)|^2 + |h_{\times}(t)|^2) \,
  \rmd t} ~.
\end{equation}
This quantity has units of s$^{1/2}$, or equivalently $\mathrm{Hz}^{-1/2}$.
In general, the root-sum-squared signal measured by a given detector,
$\hrssdet$, will be somewhat smaller.
The Monte Carlo approach taken for this analysis is to generate a set
of signals all with fixed $\hrss$ and then to add this set of signals to
the data with several discrete scale factors to evaluate different
signal amplitudes.  For a given signal morphology and $\hrss$, the
{\em efficiency} of the pipeline is the fraction of simulated signals
which are successfully recovered.

For this analysis, we do not attempt to survey the complete spectrum
of astrophysically motivated signals, but rather we use a limited
number of ad-hoc waveforms to characterize the sensitivity of the
search in terms of $\hrss$.  
Similar sensitivities may be expected for different waveforms with
similar overall properties (central frequency, bandwidth,
duration); the degree to which this is true has been
investigated in \cite{s3burst} and \cite{LVburstIa}.
The waveforms
evaluated in the present analysis are:
\begin{itemize}
\item Sine-Gaussian: sinusoid with a given frequency $f_0$ inside a Gaussian
  amplitude envelope with dimensionless width $Q$ and arrival time $t_0$:
  \begin{equation}
     h(t_0+t) =
 h_0 \sin(2\pi f_0 t) \exp\left(- \left(2\pi f_0 t\right)^2 / 2 Q^2 \right) \,.
  \end{equation}
 These are generated with linear polarization,
  with $f_0$ ranging from $70$~Hz
  to $1053$~Hz and with $Q$ equal to $3$, $8.9$, and $100$.  The
  signal consistency tests described in section~\ref{sec:consistency}
  were developed using an ensemble of sine-Gaussian signals with all
  simulated frequencies and $Q$ values.
\item Gaussian: a simple unipolar waveform with a given width $\tau$
 and linear polarization:
  \begin{equation}
     h(t_0+t) = h_0 \exp(-t^2/\tau^2) \, .
  \end{equation}
\item Band-limited white noise burst: a random signal with
  two independent polarization components
  that are white over a given frequency band, described by a
  base frequency $f_0$ and a bandwidth $\Delta f$ ({\it i.e.}\
  containing frequencies from $f_0$ to $f_0 + \Delta f$).  The signal amplitude
  has a Gaussian time envelope with a width $\tau$.
  Because these waveforms have two uncorrelated polarizations (in a
  coordinate system at some random angle), they
  provide a stringent check on the robustness of our cross-correlation test.
\end{itemize}
In all cases, we
generate each simulated signal with a random arrival direction and a
random angular relationship between the wave polarization basis and
the Earth.

\begin{figure}
\begin{center}
\begin{tabular}{cc}
\raisebox{3cm}[0pt]{(a)} &
\includegraphics[width=0.90\linewidth]{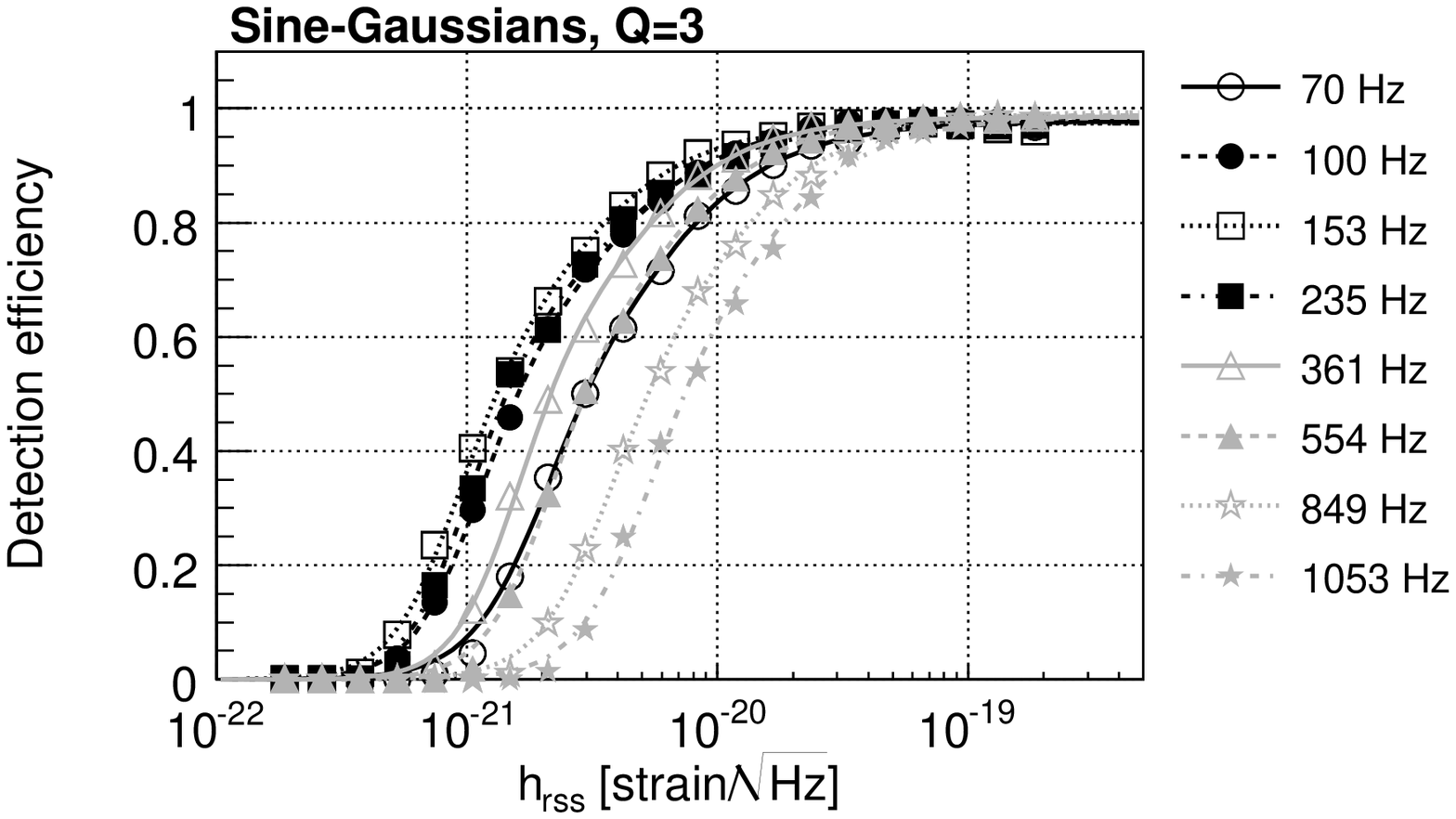} \\
\raisebox{3cm}[0pt]{(b)} &
\includegraphics[width=0.90\linewidth]{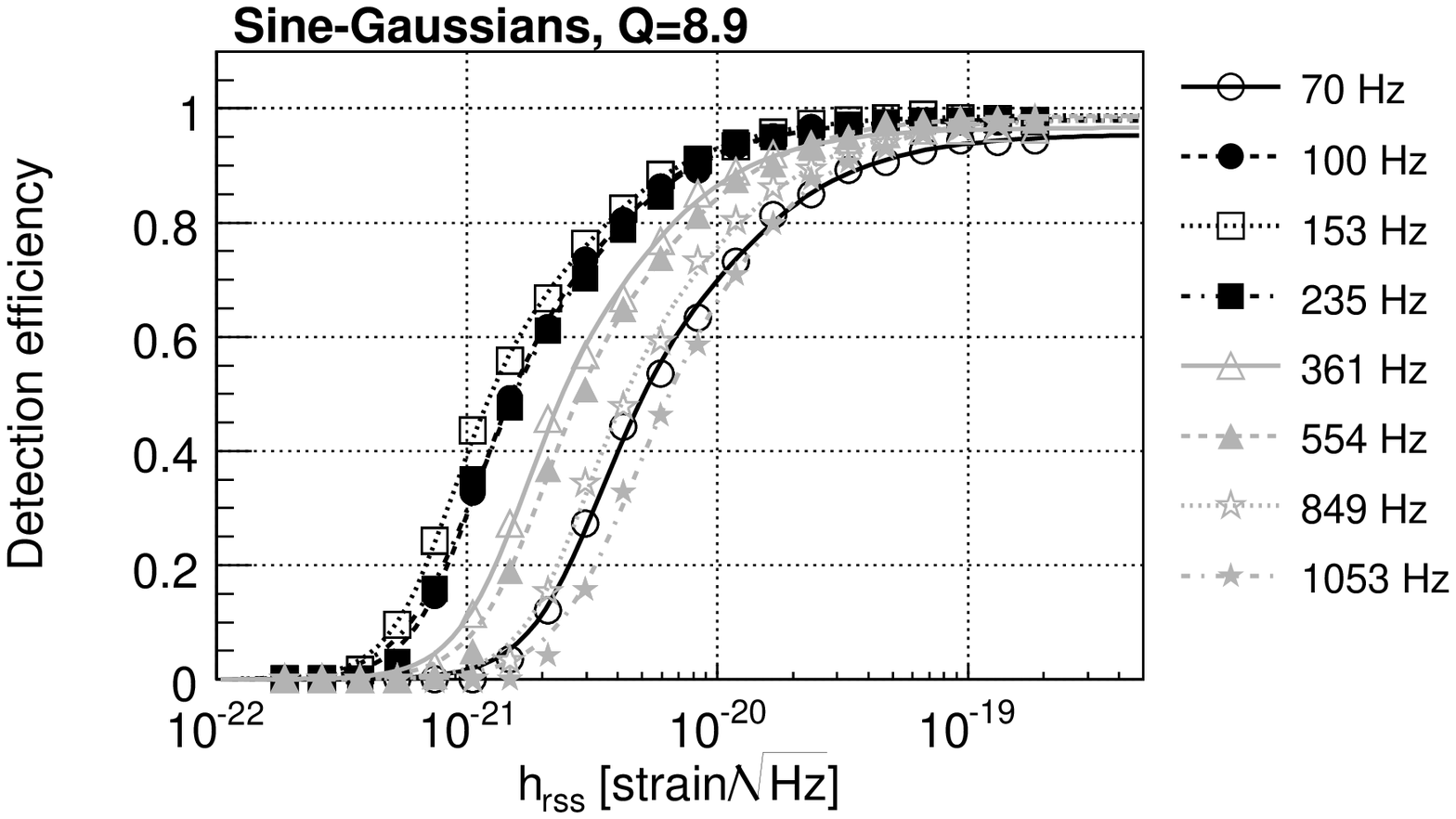} \\
\raisebox{3cm}[0pt]{(c)} &
\includegraphics[width=0.90\linewidth]{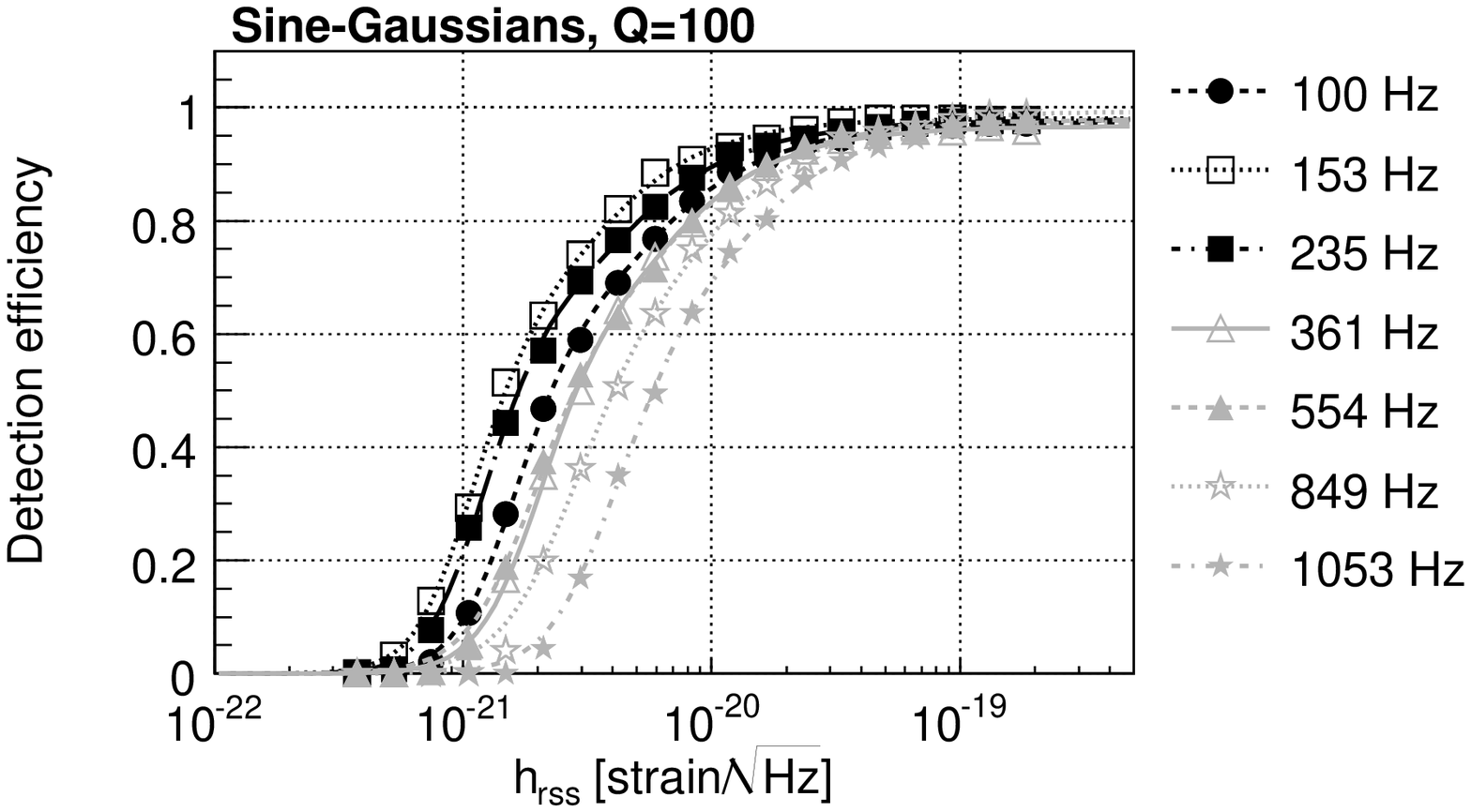} \\
\end{tabular}
\caption{Efficiency curves for simulated gravitational-wave signals:
linearly-polarized sine-Gaussian waves with
(a) $Q$=3; (b) $Q$=$8.9$; (c) $Q$=$100$.
Statistical errors are comparable to the size of the plot symbols.}
\label{fig:EffUL1}
\end{center}
\end{figure}
\begin{figure}
\begin{center}
\begin{tabular}{cc}
\raisebox{3cm}[0pt]{(a)} &
\includegraphics[width=0.90\linewidth]{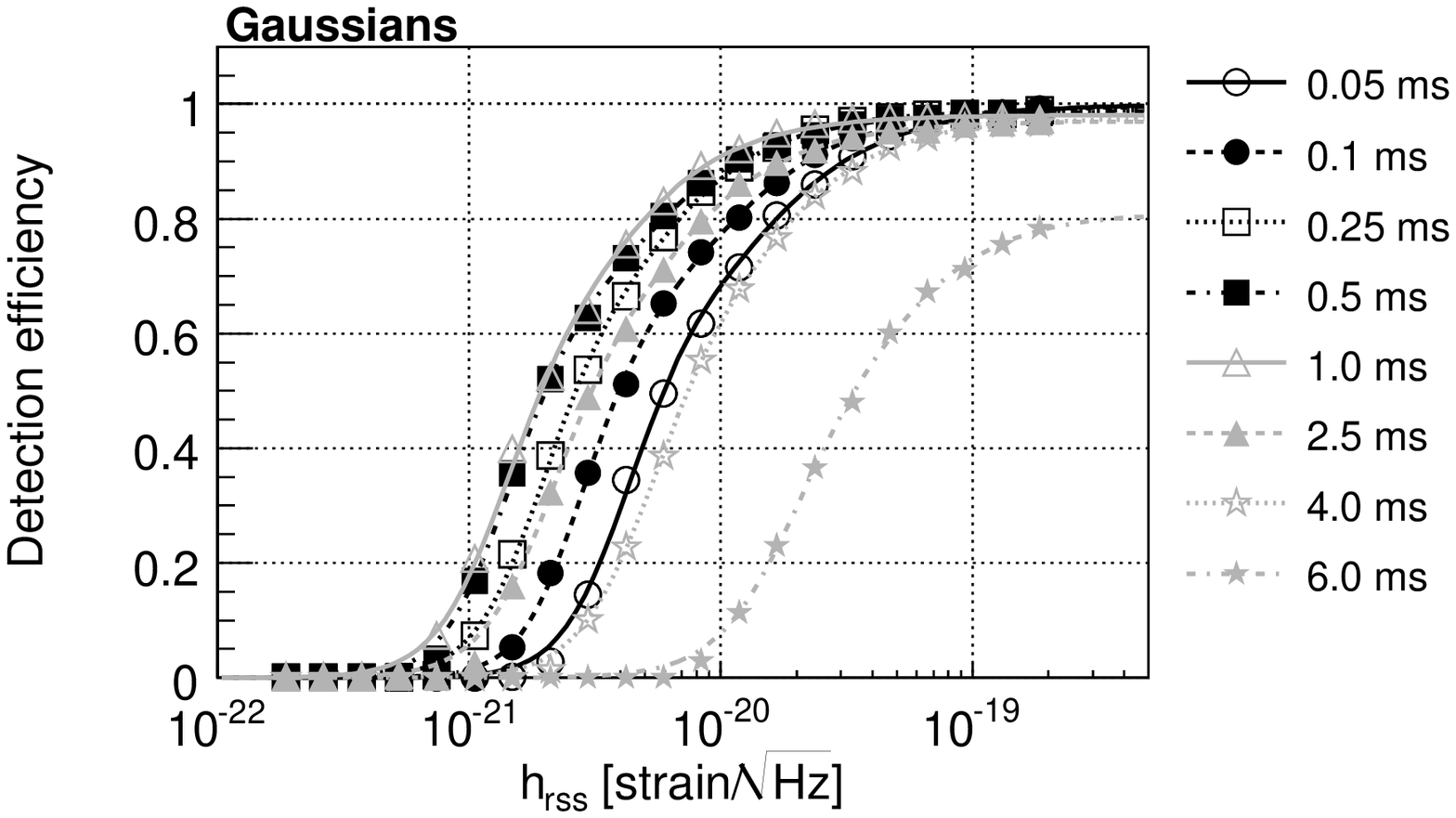} \\
\raisebox{3cm}[0pt]{(b)} &
\includegraphics[width=0.90\linewidth]{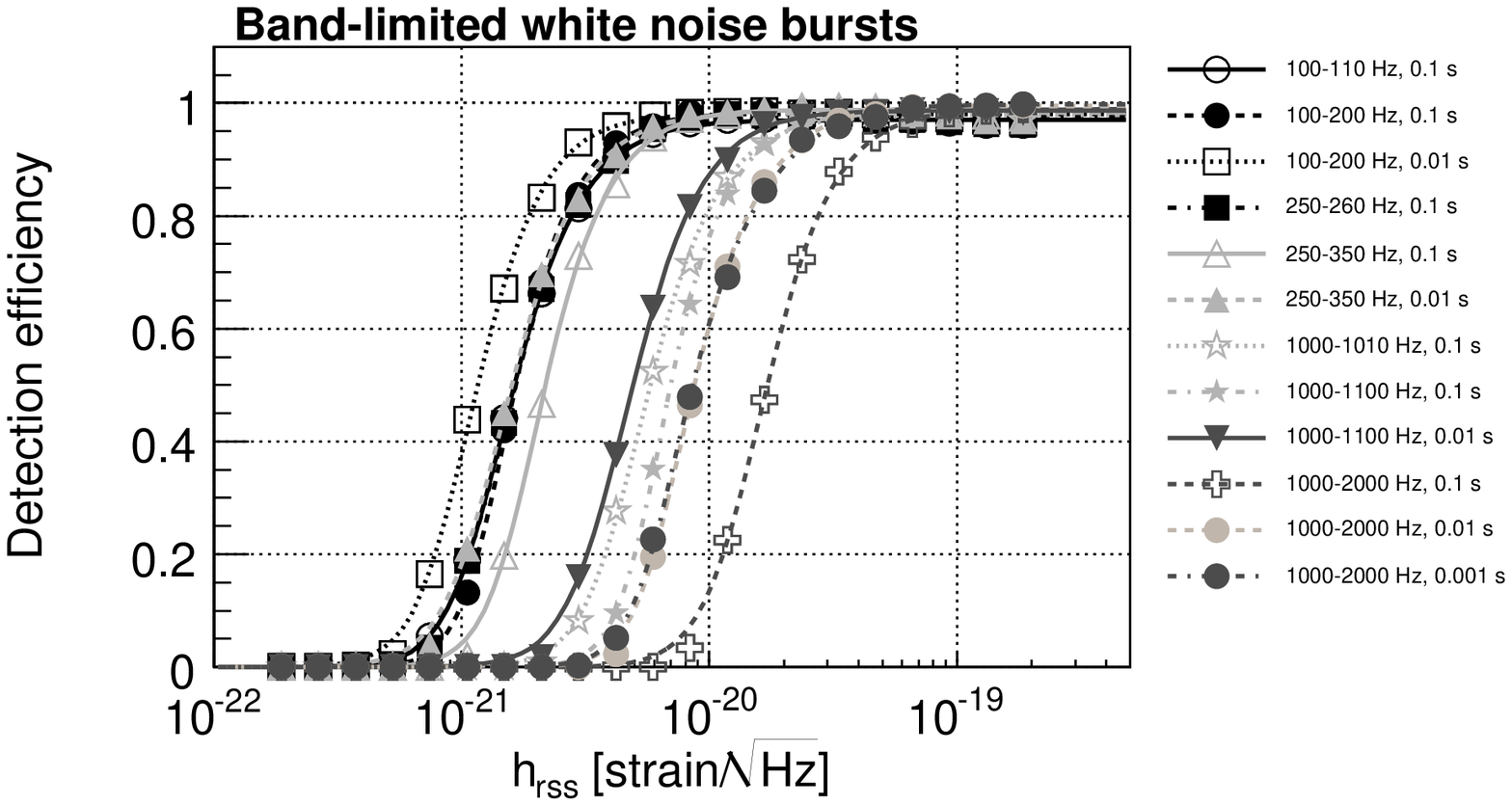} \\
\end{tabular}
\caption{Efficiency curves for simulated gravitational-wave signals:
(a) linearly-polarized Gaussian waves; (b) band-limited white-noise bursts
with two independent polarization components.
Note that four curves in the latter plot are nearly identical:
100--110~Hz, $0.1$~s; 100--200~Hz, $0.1$~s; 250--260~Hz, $0.1$~s; and
250--350~Hz, $0.01$~s.
Statistical errors are comparable to the size of the plot symbols.}
\label{fig:EffUL2}
\end{center}
\end{figure}

Figures \ref{fig:EffUL1} and \ref{fig:EffUL2}
show the measured efficiency of the analysis pipeline as a function
of root-sum-squared strain amplitude,
$\epsilon(\hrss)$, for each simulated waveform.  The
efficiency data points for each waveform are fit with a function of the
form
\begin{equation} \label{fitform}
\epsilon(\hrss) = \frac{\epsilon_\mathrm{max}}{1 +
\left(\frac{\hrss}{\hrssmid}\right)^{\alpha \left( 1+\beta
\tanh(\hrss/\hrssmid) \right) }} \, ,
\end{equation}
where 
$\epsilon_\mathrm{max}$ corresponds to the efficiency for strong
signals (normally very close to unity),
$\hrssmid$ is the $\hrss$ value corresponding to an
efficiency of $\epsilon_\mathrm{max}/2$,
$\beta$ is the parameter that describes the asymmetry of the sigmoid
(with range $-1$ to $+1$),
and $\alpha$ describes the slope.
Data points with efficiency below $0.05$ are excluded
from the fit because they do not necessarily follow the functional
form, while data points with efficiency equal to $1.0$ are excluded
because their asymmetric statistical uncertainties are not
handled properly in the chi-squared fit.
The empirical functional form in equation~\ref{fitform} has been found to fit the
remaining efficiency data points well.

Note that the Gaussian waveform with $\tau=6.0$~ms has efficiency less
than $0.8$ even for the largest simulated amplitude.  This broad
waveform, with little signal power at frequencies above 64~Hz
(the lower end of the nominal search range), is at the limit of what
the search method can detect.
For some of the other waveforms, the efficiency levels off at a value
slightly less than $1.0$ due to the application of the
auxiliary-channel vetoes, which randomly coincide in time with some of
the simulated signals.  This effect is most pronounced for the
longest-duration simulated signals due to the veto logic used in this
analysis, which rejects a trigger if there is any overlap between the
reconstructed trigger duration and a vetoed time interval.
The 70-Hz sine-Gaussian with $Q$$=$$100$ has a duration longer than
1~s and is reconstructed quite poorly; it is omitted from
figure~\ref{fig:EffUL1}c and from the following results.

\begin{figure}[bt]
\begin{center}
\includegraphics[width=0.96\linewidth]{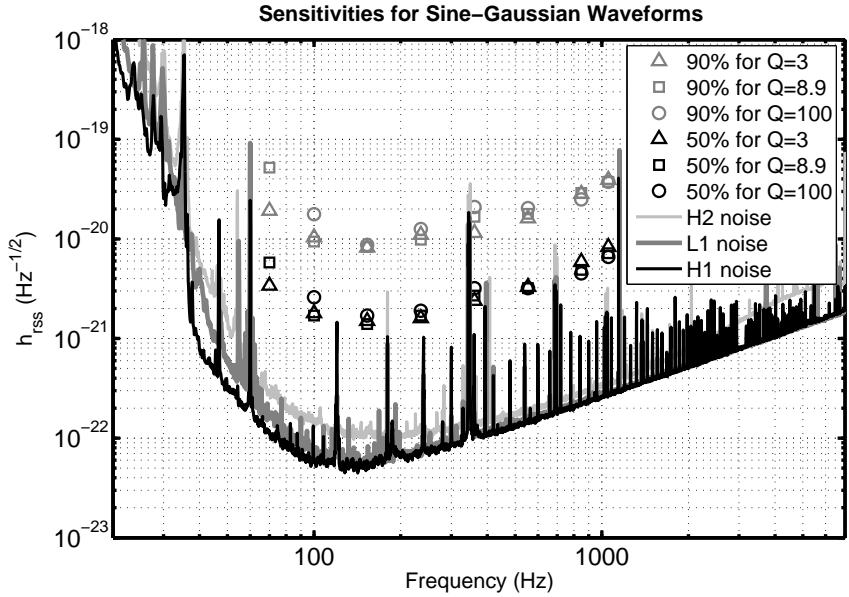}
\caption{Sensitivity of the analysis pipeline for sine-Gaussian
  waveforms as a function of frequency and $Q$.  Symbols indicate the
  $\hrss$ values corresponding to 50\% and 90\% efficiency, taken from
  table~\ref{t:effSG}.  The instrumental sensitivity curves from
  figure~\ref{fig:s4noise} are shown for comparison.}
\label{fig:sgeffplot}
\end{center}
\end{figure}

The analytic expressions of the fits are used to determine the
signal strength $\hrss$ for which efficiencies of 50\% and 90\% are
reached.  These fits are subject to statistical errors from the
limited number of simulations performed to produce the efficiency data
points.  Also, the overall amplitude scale is subject to the
uncertainty in the calibration of the interferometer response,
conservatively estimated to be 10\%~\cite{s4calib}.  We increase the
nominal fitted $\hrss$ values by the amount of these systematic uncertainties to
arrive at conservative $\hrss$ values at
efficiencies of 50\% and 90\%,
summarized in tables~\ref{t:effSG}, \ref{t:effGau}, and
\ref{t:effWNB}.
The sine-Gaussian $\hrss$ values are also displayed graphically in
figure~\ref{fig:sgeffplot}, showing how the frequency dependence
generally follows that of the instrumental noise.

\begin{table}
\caption[]{$\hrss$ values corresponding to 50\% and 90\% detection
efficiencies for simulated sine-Gaussian signals with various central
frequencies and $Q$ values.
The 70 Hz sine-Gaussian with $Q$=$100$ is not detected reliably.}
\label{t:effSG}
\begin{indented}
\item[]\begin{tabular}{ccccccc}
\br
 & \centre{6}{$\hrss$ ($10^{-21}~\mathrm{Hz}^{-1/2}$)} \\ \ns
 & \crule{6} \\
 & \centre{3}{50\% efficiency} & \centre{3}{90\% efficiency} \\ \ns
Central & \crule{3} & \crule{3} \\
frequency (Hz) &
   $Q$=$3$ & $Q$=$8.9$ & $Q$=$100$  & $Q$=$3$ & $Q$=$8.9$ & $Q$=$100$ \\
\mr
 70 &  3.4 & 5.8 & ---    & 19.2 & 52.0 & ---  \\
100 &  1.8 & 1.7 & 2.6    & 10.4 & 9.4 & 17.7  \\
153 &  1.5 & 1.4 & 1.7    &  8.2 & 8.3 & 8.7  \\
235 &  1.6 & 1.7 & 1.9    & 11.0 & 9.8 & 12.6  \\
361 &  2.4 & 2.7 & 3.2    & 11.5 & 16.7 & 20.9  \\
554 &  3.3 & 3.2 & 3.2    & 16.1 & 17.9 & 20.4  \\
849 &  5.9 & 4.9 & 4.5    & 28.4 & 28.9 & 24.9  \\
1053 & 8.3 & 7.2 & 6.6    & 39.3 & 37.5 & 37.5  \\
\br
\end{tabular}
\end{indented}
\end{table}

\begin{table}
\caption[]{$\hrss$ values corresponding to 50\% and 90\% detection
efficiencies for simulated Gaussian signals with various widths.
The waveform with $\tau$$=$$6.0$~ms does not reach an efficiency of
90\% within the range of signal amplitudes simulated.}
\label{t:effGau}
\begin{indented}
\item[]\begin{tabular}{ccc}
\br
 & \centre{2}{$\hrss$ ($10^{-21}~\mathrm{Hz}^{-1/2}$)} \\ \ns
 & \crule{2} \\
$\tau$ (ms) &  50\% efficiency & 90\% efficiency \\
\mr
$0.05$   &   6.6  &  33.9  \\
 $0.1$   &   4.4  &  25.3  \\
$0.25$   &   3.0  &  14.3  \\
 $0.5$   &   2.2  &  13.5  \\
 $1.0$   &   2.2  &  10.6  \\
 $2.5$   &   3.4  &  20.5  \\
 $4.0$   &   8.3  &  43.3  \\
 $6.0$   &  39.0  &  ---   \\
\br
\end{tabular}
\end{indented}
\end{table}

\begin{table}
\caption[]{$\hrss$ values corresponding to 50\% and 90\% detection
efficiencies for simulated ``white noise burst'' signals with various
base frequencies, bandwidths, and durations.}
\label{t:effWNB}
\begin{indented}
\item[]\begin{tabular}{ccccc}
\br
 & & & \centre{2}{$\hrss$ ($10^{-21}~\mathrm{Hz}^{-1/2}$)} \\ \ns
Base frequency & Bandwidth & Duration & \crule{2} \\
 (Hz) & (Hz) & (s) &  50\% eff.\ & 90\% eff.\ \\
\mr
 100 & 10 & $0.1$      &  1.8  &   4.7  \\
 100 & 100 & $0.1$     &  1.9  &   4.1  \\
 100 & 100 & $0.01$    &  1.3  &   2.9  \\
 250 & 10 & $0.1$      &  1.8  &   4.5  \\
 250 & 100 & $0.1$     &  2.4  &   5.4  \\
 250 & 100 & $0.01$    &  1.8  &   4.3  \\
 1000 & 10 & $0.1$     &  6.5  &  15.8  \\
 1000 & 100 & $0.1$    &  7.9  &  16.7  \\
 1000 & 100 & $0.01$   &  5.5  &  12.7  \\
 1000 & 1000 & $0.1$   & 19.2  &  42.6  \\
 1000 & 1000 & $0.01$  &  9.7  &  22.3  \\
 1000 & 1000 & $0.001$ &  9.5  &  23.7  \\
\br
\end{tabular}
\end{indented}
\end{table}

\begin{figure}[bt]
\begin{center}
$
\begin{tabular}{c}
\includegraphics[width=0.96\linewidth]{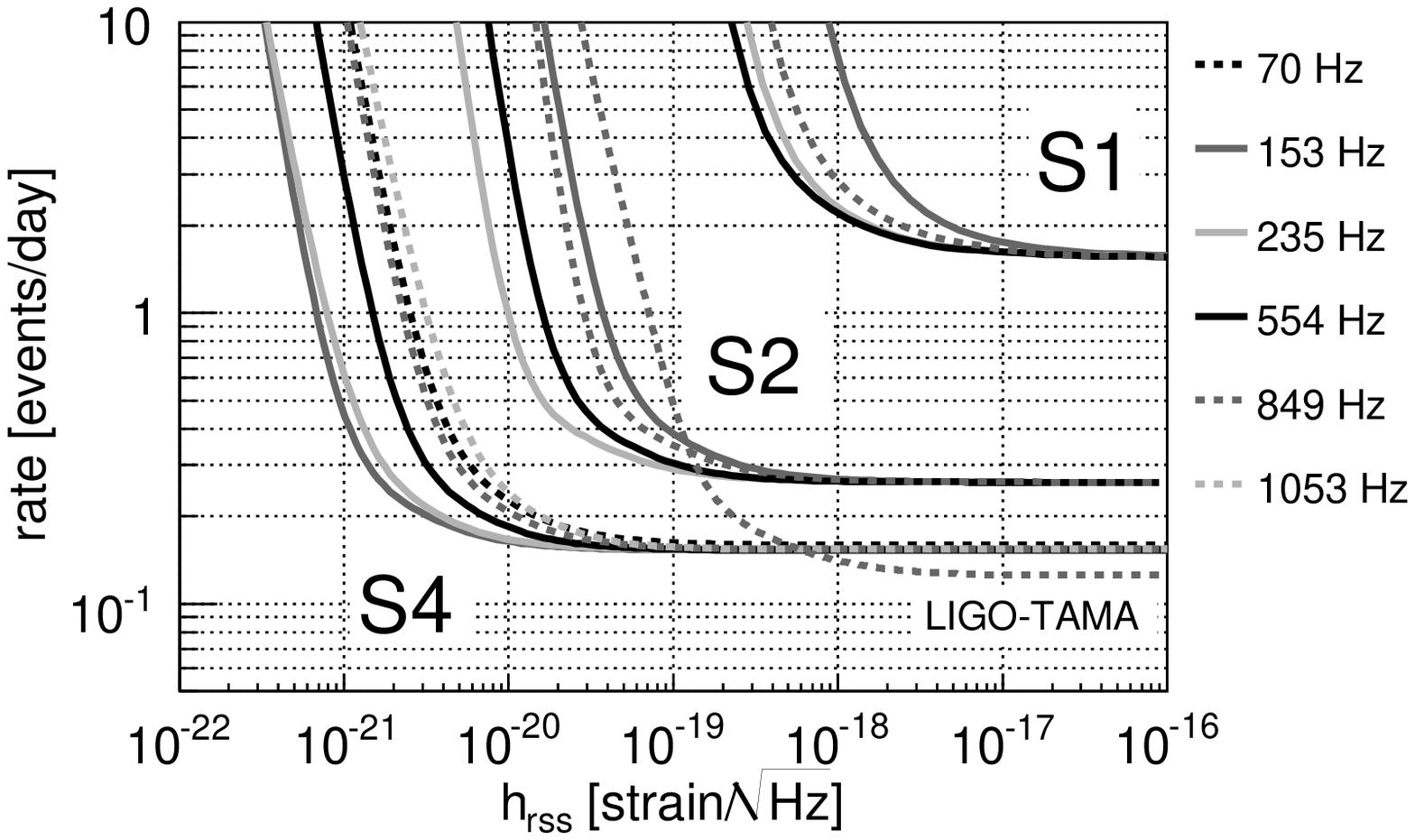} \\
\mbox{(a) Sine-Gaussians with $Q$$=$$8.9$} \\ 
\includegraphics[width=0.96\linewidth]{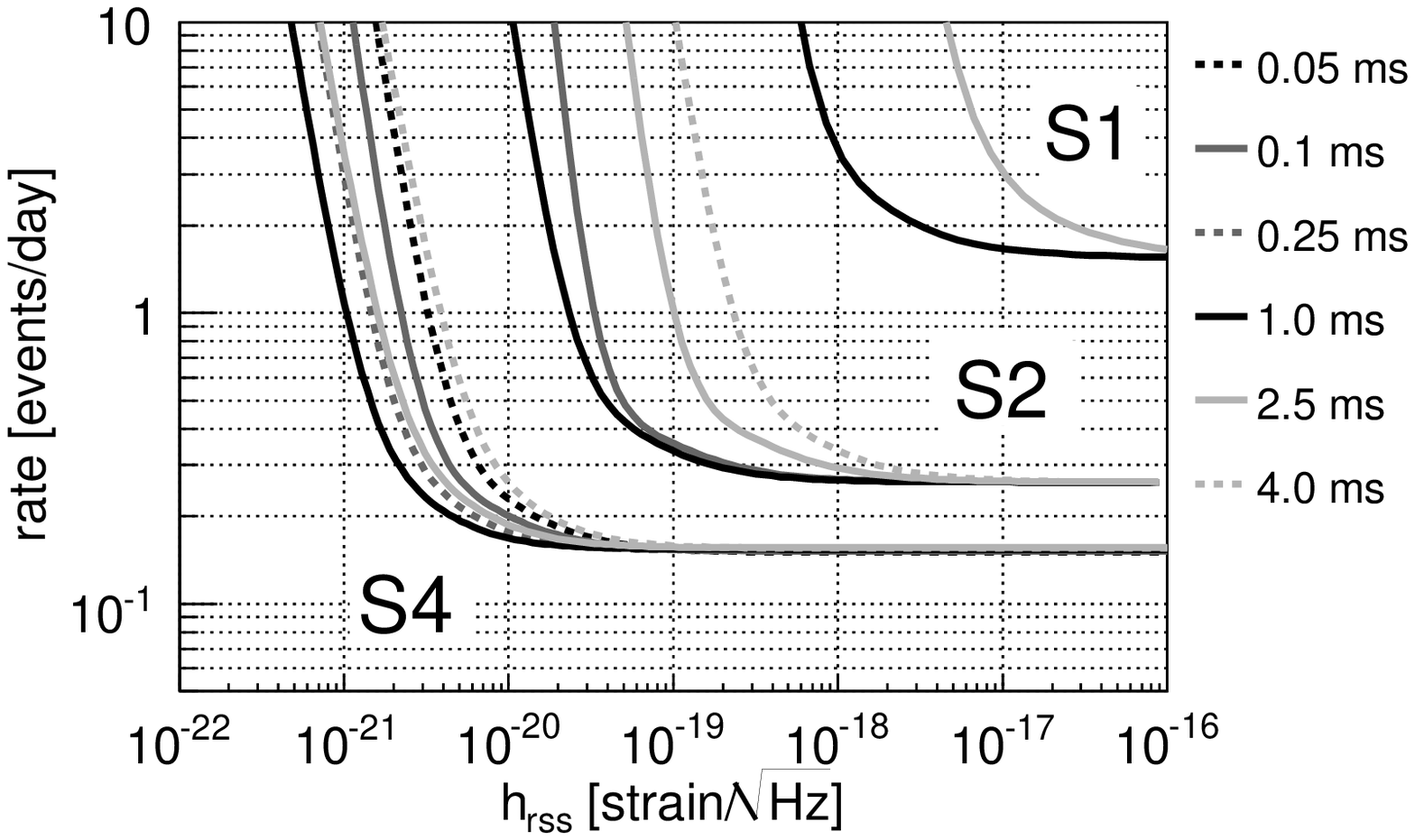} \\
\mbox{(b) Gaussians} \\
\end{tabular}
$
\caption{Exclusion diagrams (rate limit at 90\% confidence level, as
a function of signal amplitude) for (a) sine-Gaussian and (b) Gaussian
simulated waveforms for this S4 analysis compared to the S1 and S2
analyses (the S3 analysis did not state a rate limit).
These curves incorporate conservative systematic uncertainties from
the fits to the efficiency curves and from the interferometer response
calibration.  The 849~Hz curve labeled ``LIGO-TAMA'' is from the joint
burst search using LIGO S2 with TAMA DT8 data~\cite{S2LIGOTAMA}, which
included data subsets with different combinations of operating
detectors with a total observation time of $19.7$~days and thereby
achieved a lower rate limit.  The $\hrss$ sensitivity of the LIGO-TAMA
search was nearly constant for sine-Gaussians over the frequency range
$700$--$1600$~Hz.}

\label{fig:S1S2S4}
\end{center}
\end{figure}

Event rate limits as a function of waveform type and signal
amplitude can be represented by an ``exclusion diagram''.
Each curve in an exclusion diagram indicates what the rate limit
would be for a population of signals with a fixed $\hrss$, as a
function of $\hrss$.  The curves in figure~\ref{fig:S1S2S4}
illustrate, using selected sine-Gaussian and Gaussian waveforms that
were also considered in the S1 and S2 analyses, that the
amplitude sensitivities achieved by this S4 analysis are at least
an order of magnitude better than the sensitivities achieved by the S2
analysis.  For instance,
the 50\% efficiency $\hrss$ value for 235 Hz sine-Gaussians with
$Q$$=$$8.9$ is $1.5 \times 10^{-20}~\mathrm{Hz}^{-1/2}$ for S2 and
$1.7 \times 10^{-21}~\mathrm{Hz}^{-1/2}$ for S4.
(Exclusion curves were not generated
for the S3 analysis, but the S3 sensitivity
was $9 \times 10^{-21}~\mathrm{Hz}^{-1/2}$ for this particular waveform.)
The improvement is greatest for lower-frequency sine-Gaussians and
for the widest Gaussians, due to the reduced low-frequency detector
noise and the explicit extension of the search band down to $64$~Hz.

\section{Astrophysical reach estimates}  \label{sec:reach}

In order to set an astrophysical scale to the sensitivity achieved
by this search,
we can ask what amount of mass converted into gravitational-wave
burst energy at a given distance would be strong
enough to be detected by the search pipeline with 50\% efficiency.
We start with the expression for the instantaneous
energy flux emitted by a gravitational wave source in the two independent
polarizations $h_{+}(t)$ and $h_{\times}(t)$~\cite{astropheno},
\begin{equation}\label{eq:flux}
\frac{\rmd^2E_\mathrm{GW}}{\rmd A\,\rmd t}=\frac{1}{16\pi} \frac{c^3}{G}\left<(\dot{h}_{+})^2+(\dot{h}_{\times})^2\right> \, ,
\end{equation}
and follow the derivations in~\cite{riles}.
Plausible astrophysical sources will, in general, emit gravitational
waves anisotropically, but here we will assume isotropic emission in
order to get simple order-of-magnitude estimates.
The above formula, when integrated over the signal duration
and over the area of a sphere at radius $r$
(assumed not to be at a cosmological distance), yields the total energy
emitted in gravitational waves for a given signal waveform.
For the case of a sine-Gaussian with frequency $f_0$ and $Q \gg 1$,
we find
\begin{equation}\label{eq:SGenergy}
E_\mathrm{GW}= \frac{r^2 c^3}{4 G} (2\pi f_0)^2 \hrss^2 \, .
\end{equation}
Taking the waveform for which we have the best $\hrss$ sensitivity,
a 153 Hz sine-Gaussian with $Q$=$8.9$,
and assuming a typical Galactic source distance of 10~kpc, the above formula
relates the 50\%-efficiency $\hrss=1.4\times10^{-21}~\mathrm{Hz}^{-1/2}$
to $10^{-7}$ solar mass equivalent emission into a gravitational-wave
burst from this hypothetical source and under the given
assumptions.
For a source in the Virgo galaxy cluster, approximately 16~Mpc away,
the same $\hrss$ would be produced by an energy emission of
roughly $0.25\,M_\odot c^2$ in a burst with this highly favourable waveform.

We can draw more specific conclusions about detectability for models
of astrophysical sources which predict the absolute energy and
waveform emitted.  Here we consider the core-collapse supernova
simulations of Ott {\it et al.}~\cite{OBDL} and a binary black hole
merger waveform calculated by the Goddard numerical relativity
group~\cite{Bakeretal} (as a representative example of the similar
merger waveforms obtained by several groups).  While the Monte Carlo
sensitivity studies in section~\ref{sec:sensitivity} did not include
these particular waveforms, we can relate the modeled waveforms to
qualitatively similar waveforms that {\em were} included in the
Monte Carlo study and thus infer the approximate sensitivity of the
search pipeline for these astrophysical models.

Ott {\it et al.}\ simulated core collapse for three progenitor models
and calculated the resulting gravitational wave emission, which was
dominated by oscillations of the protoneutron star core driven by
accretion~\cite{OBDL}.  Their s11WW model, based on a non-spinning
11-$M_\odot$ progenitor, produced a total gravitational-wave energy
emission of $1.6 \times 10^{-8}\,M_\odot c^2$ with a characteristic
frequency of $\sim$$654$~Hz and duration of several hundred
milliseconds.  If this were a sine-Gaussian, it would have a $Q$ of
several hundred; table~\ref{t:effSG} shows that our sensitivity does
not depend strongly on $Q$, so we might expect 50\% efficiency for a
signal at this frequency with $\hrss$ of
$\sim$$3.7 \times 10^{-21}~\mathrm{Hz}^{-1/2}$.
However, the signal is not monochromatic, and its increased
time-frequency volume may degrade the sensitivity by up to a
factor of $\sim$$2$.  Using this $E_\mathrm{GW}$ and
$\hrss \approx 7 \times 10^{-21}~\mathrm{Hz}^{-1/2}$
in equation~\ref{eq:SGenergy}, we find that our
search has an approximate ``reach'' (distance for which the signal
would be detected with 50\% efficiency by the analysis pipeline)
of $\sim$$0.2$~kpc for this model.
The m15b6 model, based on a spinning 15-$M_\odot$ progenitor, yields a
very similar waveform and essentially the same reach.  The s25WW
model, based on a 25-$M_\odot$ progenitor, was found to emit vastly
more energy in gravitational waves, $8.2 \times 10^{-5}\,M_\odot c^2$,
but with a higher characteristic frequency of $\sim$$937$~Hz.  With
respect to the Monte Carlo results in section~\ref{sec:sensitivity},
we may consider this similar to a high-Q sine-Gaussian, yielding
$\hrss \approx 5.5 \times 10^{-21}~\mathrm{Hz}^{-1/2}$,
or to a white noise burst with a
bandwidth of $\sim$$100$~Hz and a duration of $> 0.1$~s, yielding
$\hrss \approx 8 \times 10^{-21}~\mathrm{Hz}^{-1/2}$.
Using the latter, we deduce an
approximate reach of $8$~kpc for this model.

A pair of merging black holes emits gravitational waves with very high
efficiency; for instance, numerical evolutions of equal-mass systems
without spin have found the radiated energy from the merger and
subsequent ringdown to be $3.5$\% or more of the total mass of the
system~\cite{Bakeretal}.  From figure~8 of that paper, the
frequency of the signal at the moment of peak amplitude is seen to be
\begin{equation}
  f_{\mathrm{peak}} \approx \frac{0.46}{2 \pi M_f}
                  \approx \frac{15~\mathrm{kHz}}{(M_f/M_\odot)} \, ,
\end{equation}
where $M_f$ is the final mass of the system.  Very roughly, we can
consider the merger+ringdown waveform to be similar to a sine-Gaussian
with central frequency $f_{\mathrm{peak}}$ and $Q \approx 2$ for
purposes of estimating the reach of this search pipeline for binary
black hole mergers.  (Future analyses will include Monte Carlo
efficiency studies using complete inspiral-merger-ringdown waveforms.)
Thus, a binary system of two 10-$M_\odot$ black holes ({\it i.e.} $M_f
\approx 20 \, M_\odot$) has $f_\mathrm{peak} \approx 750$~Hz, and from
table~\ref{t:effSG} we can estimate the $\hrss$ sensitivity to be
$\sim$$5.5 \times 10^{-21}~\mathrm{Hz}^{-1/2}$.
Using $E_{\mathrm GW} = 0.035 \, M_f c^2$,
we conclude that the reach for such a system is roughly $1.4$~Mpc.
Similarly, a binary system with $M_f = 100 \, M_\odot$ has
$f_\mathrm{peak} \approx 150$~Hz, a sensitivity of $\sim$$1.5 \times
10^{-21}~\mathrm{Hz}^{-1/2}$, and a resulting reach of roughly $60$~Mpc.

\section{Discussion}  \label{sec:discussion}

The search reported in this paper represents the most sensitive search
to date for gravitational-wave bursts in terms of strain amplitude,
reaching $\hrss$ values below $10^{-20}~\mathrm{Hz}^{-1/2}$,
and covers a broad frequency range, $64$--$1600$~Hz, with a live
observation time of $15.5$ days.

Comparisons with previous LIGO~\cite{s1burst,s2burst} and
LIGO-TAMA~\cite{S2LIGOTAMA} searches have already been shown
graphically in figure~\ref{fig:S1S2S4}.
The LIGO-TAMA search targeted millisecond-duration
signals with frequency content in the $700$--$2000$~Hz frequency regime
({\it i.e.},
partially overlapping the present search) and had a detection 
efficiency of at least 50\% (90\%) for signals with
$\hrss$ greater than $\sim 2\times 10^{-19}~\mathrm{Hz}^{-1/2}$
($10^{-18}~\mathrm{Hz}^{-1/2}$).
Among other searches with broad-band interferometric
detectors~\cite{ref:UGMPQ,ref:Forward,ref:TAMADT9}, the most recent
one by the TAMA collaboration reported an upper limit of $0.49$ events
per day at the 90\% confidence level based on an analysis of $8.1$ days
of the TAMA300 instrument's
ninth data taking run (DT9) in 2003--04.
The best sensitivity of this TAMA search
was achieved when looking for narrow-band signals
at TAMA's best operating frequency, around $1300$~Hz, and it was at
$\hrss\approx 10^{-18}~\mathrm{Hz}^{-1/2}$ for 50\% detection
efficiency~\cite{ref:TAMADT9}.
Although we did not measure the sensitivity of the S4 LIGO
search with narrow-band signals at $1300$~Hz, LIGO's noise
at that frequency range varies slowly enough so that we do not expect
it to be significantly worse than the sensitivity for $1053$~Hz sine-Gaussian
signals described in section~\ref{sec:sensitivity},
which stands at about $7 \times 10^{-21}~\mathrm{Hz}^{-1/2}$.

Comparisons with results from resonant mass detectors
were detailed
in our previous publications~\cite{s1burst,s2burst}. The upper limit
of $\sim 4\times 10^{-3}$ events per day
at the 95\% confidence level on the rate of gravitational wave bursts
set by the IGEC consortium of five resonant mass detectors still
represents the most stringent rate limit for $\hrss$ signal strengths of order
$10^{-18}~\mathrm{Hz}^{-1/2}$ and above~\cite{IGEC}. This upper limit
quickly falls off and
becomes inapplicable to signals weaker than $10^{-19}~\mathrm{Hz}^{-1/2}$
(see figure 14 in~\cite{s2burst}.)
Furthermore, with the improvement in our search sensitivity,
the signal strength of the events corresponding to the
slight excess seen by the EXPLORER and NAUTILUS resonant mass detectors
in their 2001 data~\cite{rome2001} falls well above the 90\%
sensitivity of our current S4 search:
as described in~\cite{s2burst}, the optimal orientation
signal strength of these events assuming a Gaussian morphology
with $\tau$=0.1~ms corresponds to a
$\hrss$ of $1.9\times 10^{-19}~\mathrm{Hz}^{-1/2}$.
For such Gaussians our S4 search all-sky 90\% sensitivity
is $2.5 \times 10^{-20}~\mathrm{Hz}^{-1/2}$ (see Table~\ref{t:effGau})
and when accounting for {\it optimal} orientation,
this improves by roughly a factor of 3,
to $9.3 \times 10^{-21}~\mathrm{Hz}^{-1/2}$.
The rate of the EXPLORER and NAUTILUS events was of order 
200 events/year (or 0.55 events per day)~\cite{rome2001,coccia2004}.
A steady flux of gravitational-wave bursts at this rate is excluded
by our present measurement at the 99.9\% confidence level.
Finally, in more recent running of the EXPLORER and NAUTILUS detectors,
an analysis of $149$ days of data collected in 2003
set an upper limit of $0.02$
events per day at the 95\% confidence level and with a 
$\hrss$ sensitivity of
$\sim 2\times 10^{-19}~\mathrm{Hz}^{-1/2}$~\cite{rome2003}.

The S5 science run, which began in November 2005 and is
expected to continue until late 2007, has a goal of
collecting a full year of coincident LIGO science-mode data.  Searches for
gravitational-wave bursts using S5 data are already underway and will
be capable of detecting any sufficiently strong signals which arrive
during that time, or else placing an upper limit on the rate of such
signals on the order of a few per year.
Furthermore, the detector
noise during the S5 run has reached the design goals for the current
LIGO interferometers, and so the amplitude sensitivity of S5 burst
searches is expected to be roughly a factor of two better than the
sensitivity of this S4 search.

Another direction being pursued with the S5 data is to make
appropriate use of different detector network configurations.  In
addition to the approach used in the S4 analysis reported here, which
requires a signal to appear with excess power in a time-frequency
map in all three LIGO interferometers, data
from two-detector combinations is also being analyzed to maximize
the total observation time.  Furthermore, using LIGO data together
with simultaneous data from other interferometers can significantly
improve confidence in a signal candidate and allow more properties of
the signal to be deduced.  The GEO 600
interferometer has joined the S5 run for full-time observing in May
2006, and we look forward to the time when VIRGO begins operating with
sensitivity comparable to the similarly-sized LIGO interferometers.
Members of the LSC are currently implementing coherent network analysis
methods using maximum likelihood approaches for optimal detection of
arbitrary burst signal (see, for example, \cite{ConstraintL})
and for robust signal consistency tests~\cite{WenSchutz,NetConsist}.
Such methods will make the best use of the data collected from the
global network of detectors to search for gravitational-wave bursts.

\section*{Acknowledgments}

The authors gratefully acknowledge the support of the United States
National Science Foundation for the construction and operation of the
LIGO Laboratory and the Science and Technology Facilities Council of the
United Kingdom, the Max-Planck-Society, and the State of
Niedersachsen/Germany for support of the construction and operation of
the GEO600 detector. The authors also gratefully acknowledge the support
of the research by these agencies and by the Australian Research Council,
the Council of Scientific and Industrial Research of India, the Istituto
Nazionale di Fisica Nucleare of Italy, the Spanish Ministerio de
Educaci\'on y Ciencia, the Conselleria d'Economia, Hisenda i Innovaci\'o of
the Govern de les Illes Balears, the Scottish Funding Council, the
Scottish Universities Physics Alliance, The National Aeronautics and
Space Administration, the Carnegie Trust, the Leverhulme Trust, the David
and Lucile Packard Foundation, the Research Corporation, and the Alfred
P. Sloan Foundation.
This document has been assigned LIGO Laboratory document number \dccnumber.

\end{document}